\documentclass[smallextended]{svjour3}       % onecolumn (second format)
\smartqed  % flush right qed marks, e.g. at end of proof
\usepackage{graphicx,epsfig}
\begin{document}

\title{Temperature-field phase diagrams of one-way quantum work deficit in two-qubit
XXZ spin systems}

\author{
M.A.Yurischev
}

\institute{
M.~A.~Yurischev
\at
%Quantum Correlation(s) Research Center,\\
Institute of Problems of Chemical Physics, Russian Academy of Sciences,
Chernogolovka 142432, Moscow Region, Russia\\     
\email{yur@itp.ac.ru} }

\date{Received:}

\titlerunning{
Temperature-field phase diagrams of one-way quantum work deficit
}
\maketitle

\begin{abstract}
The spin-1/2 XXZ chain in a uniform magnetic field at thermal equilibrium is
considered.
For this model, we give a complete classification of all qualitatively different phase
diagrams for the one-way quantum work (information) deficit.
The diagrams can contain regions (phases, fractions) with both stationary and
variable (state-dependent) angles of optimal measurement.
We found cases of phase diagrams in which the sizes of regions with the variable
optimal measurement angle are large and perhaps such regions can be detected
experimentally.
We also established a relationship between the behavior of optimal measurement angles
near the boundaries separated different regions and Landau's theory of phase
transitions of the second and first kind.
\end{abstract}

%\PACS{03.65.Aa \and 03.65.Ta \and 03.67. -a \and 89.70.Cf}
% \and 42.50. -p}

%======================================================================
\section{Introduction}
\label{sect:Intro}
Quantum correlations have foundational interest and are an essential resource for
quantum information and computational technologies.
There are many measures of quantum correlations.
One of the most important places among them belongs to the quantum discord and one-way
quantum work deficit \cite{MBCPV12,Str15,ABC16,FPA17,BDSRSS18}.
It is significant that the concept of these correlations, as opposed to others,
is introduced through {\em measurements}\/ and therefore is consistent with the
fundamental requirement:
``Unless a thing can be defined by measurement, it has no place in a theory''
\cite{FLS64}.
These words repeat the idea of Heisenberg's 1925 paper \cite{H25}.
(In this connection, see, e.g., \cite{W14}.)

The quantum discord ${\cal Q}$ for a bipartite system $AB$ is defined as the minimum
difference between the quantum generalizations of symmetric ($I$) and asymmetric
looking ($J$) versions of the classical mutual information:
${\cal Q}=\min_{\{{\rm\Pi}_k\}}(I-J)$, where $\{{\rm\Pi}_k\}$ is the measurement
performed on one of the two subsystems (say, on $B$) \cite{Z00,OZ02,HV01,V17}.
One can also rewrite this definition in equivalent form as the minimum difference
between the quantum conditional entropy after a local measurement (Zurek's definition
\cite{Z00}, see also \cite{L73}), $S(A|B_{\{{\rm\Pi}_k\}})$, and quantum
pseudo-conditional entropy $S(A|B)=S(A,B)-S(B)$; that is,
${\cal Q}=\min_{\{{\rm\Pi}_k\}}S(A|B_{\{{\rm\Pi}_k\}})-S(A|B)$.
Quantum discord equals zero for the classically correlated states and coincides the
quantum entanglement for the pure states.

The quantum work deficit is a measure of quantum correlation originated from the
thermodynamics view point.
In general, it is defined as the minimum difference between the thermodynamic work
${\cal W}$
which can be extracted from a heat bath using operations on the entire quantum system
and the largest amount of work $W$ drawn from the same heat bath by manipulating only
the local parts of composite system;
in other words, the work deficit $\rm\Delta$ is the amount of potential work which
cannot be extracted under local operations because of quantum correlations
\cite{OHHH02,HHHHOSS03,HHHOSSS05}.
Exist several forms of deficit depending on the type of communication allowed
between parts $A$ (Alice) and $B$ (Bob).
If the bipartite state $\rho_{AB}$ is shared by Alice and Bob and, for example, Bob
performs a single von Neumann measurement on his local subsystem and using classical
communication sends the resulting state to Alice that extracts the maximum amount of
work $W$ from the new entire state then the dimensionless quantity
${\rm\Delta}=\min_{\{{\rm\Pi_k}\}}({\cal W}-W)/T$ is called the one-way quantum
work deficit; $T$ is the temperature (in energy units) of the common bath.

In spite of quite different conceptual sources, the one-way deficit and discord
coincide, contrary to the entanglement, in considerably more general cases.
They are identical for the Bell-diagonal (and hence Werner) states and even for the
two-qubit X states with one zero Bloch vector if the local measurement is performed
on a qubit with the vanished Bloch vector \cite{YF16}.
On the other hand, the one-way quantum deficit and discord exhibit, generally
speaking, different quantitative and even qualitative behavior in more general cases.
As a result, we now do not know what value of quantum correlation is correct.
This situation is similar to the following:
``'a moo must be three goos', when nobody knows what a moo or a goo is'' \cite{F65}.

Two following key properties play the crucial roll in our approach.
First, for the X quantum states, the optimization of both quantum discord and one-way
deficit over the projectors $\{{\rm\Pi}_k(\theta,\phi)\}$ can be worked out exactly
over the azimuthal angle $\phi$ and therefore only one optimization procedure, in the
polar angle $\theta\in[0,\pi/2]$, remains relevant
\cite{LXSW10,CRC10,VR12,YWF16,Y18,Y19}.
Thus, the non-optimized discord and one-way deficit are the functions of one variable
only: $Q=Q(\theta)$ and $\Delta=\Delta(\theta)$, respectively.
The second remarkable property is that the first derivatives
of named functions with respect to the argument $\theta$ equal zero at both
endpoints $\theta=0$ and $\pi/2$ by any choice of state parameters.
This is proved by direct calculations.

The optimal measurement angle for the Bell-diagonal states is achieved always
on the border of the interval $[0,\pi/2]$, i.e., at the limiting points 0 or $\pi/2$.
However, for the more general X states, the optimal measurement angle can also happen
{\em inside} the open interval $(0,\pi/2)$ \cite{LMXW11,CZYYO11,H13}.
In turn, this can lead to the existence of new branch (phase, fraction) of the
piecewise-defined quantum correlation function, namely, the phase with the interior,
state-dependent (variable) optimal measurement angle:
$Q_{\theta^*}$ for the quantum discord \cite{Y14,Y14a,Y15,Y17} and
$\Delta_{\vartheta}$ for the one-way quantum work deficit \cite{YWF16,Y18,Y19}.

An observation made by the author \cite{Y14,Y14a,Y15} is that the quantum discord
function $Q(\theta)$ is {\em unimodal} in the open interval $(0,\pi/2)$.
Then from topological reasoning it follows that the boundaries between the different
fractions (phases) can be found from conditions of vanishing the second derivatives
with respect to the angle $\theta$ at the endpoints:
$Q^{\prime\prime}(0)=0$ and $Q^{\prime\prime}(\pi/2)=0$.
These equations are confirmed now for various subclasses of X states \cite{Y15,Y17}.
The same is valid for the one-way deficit function.
Moreover, the measurement-dependent deficit $\Delta(\theta)$ can also exhibit in some
cases the {\em bimodal} behavior, i.e., it can have two interior extrama: one minimum
and one maximum.
As a result, this can lead to the new mechanism of formation of the boundary between
phases, namely, via finite jumps of optimal measured angle from the
endpoint to the interior minimum or vice versa \cite{Y18,Y19}.

This paper is devoted to the systematic classification of all possible
temperature-field phase diagrams for the one-way quantum work deficit in spin-1/2 XXZ
dimers.
In Sect.~\ref{sec:Model}, the required equations are given for constructing the phase
diagrams.
Then, in Sect.~\ref{sec:PhDs} we solve the corresponding equations, graphically
present a complete set of phase diagrams for all qualitatively distinguishable cases,
and describe their properties in detail.
Finally, in Sect.~\ref{sec:Concl}, our main results are summarized and some possible
avenues for the future researches are given.

%======================================================================
\section{
Formalism
}
\label{sec:Model}
We will study the structures of piecewise-defined quantum correlation function,
namely, the one-way quantum deficit, for the two-site spin-1/2 XXZ chain under a
uniform external field $B$ at the temperature $T$.

%----------------------------------------------------------------------
\subsection{Hamiltonian, density matrix, and pre-measurement entropy
}
\label{subsect:Ham}
Let us consider a composite system $AB$ with the Hamiltonian
\begin{equation}
   \label{eq:H}
   {\cal H}=-\frac{1}{2}[J(\sigma_x^A\sigma_x^B+\sigma_y^A\sigma_y^B)
	 +J_z\sigma_z^A\sigma_z^B]
	 -\frac{1}{2}B(\sigma_z^A+\sigma_z^B),
\end{equation}
where $J$ and $J_z$ are the coupling constants and
$\sigma_x^{A,B}$, $\sigma_y^{A,B}$, and $\sigma_z^{A,B}$ the Pauli spin matrices
at sites $A$ and $B$, respectively.

The energy levels of this Hamiltonian equal
\begin{equation}
   \label{eq:Ei}
   E_{1,2}=-\frac{1}{2}J_z\pm B,\qquad E_{3,4}=\frac{1}{2}J_z\pm J
\end{equation}
and consequently the partition function $Z=\sum_i\exp(-E_i/T)$ is given as
\begin{equation}
   \label{eq:Z}
   Z=2\biggl(e^{J_z/2T}\cosh\frac{B}{T}+e^{-J_z/2T}\cosh\frac{J}{T}\biggl).
\end{equation}
The function $Z(T,B)$ is invariant under the replacements both $J\to-J$ and $B\to-B$.

The Helmholtz free energy equals ${\cal F}(T,B)=-T\ln Z$ and hence the ordinary
thermodynamic entropy $S=-\partial {\cal F}/\partial T$ is expressed for the system
(\ref{eq:H}) by equation
\begin{eqnarray}
   \label{eq:S_TB}
   &&S(T,B)=
	 -\frac{1}{Z}\biggl[\frac{B+J_z/2}{T}\exp\left(\frac{B+J_z/2}{T}\right)
	 -\frac{B-J_z/2}{T}\exp\left(-\frac{B-J_z/2}{T}\right)
	 \nonumber\\
   &&+\frac{J-J_z/2}{T}\exp\left(\frac{J-J_z/2}{T}\right)
	 -\frac{J+J_z/2}{T}\exp\left(-\frac{J+J_z/2}{T}\right)\biggr]+\ln Z.
\end{eqnarray}

The Gibbs density matrix, $\rho_{AB}=\exp(-{\cal H}/T)/Z$, has the structure
\begin{equation}
   \label{eq:rhoAB}
   \rho_{AB}
	 =\left(
      \begin{array}{cccc}
      a&&&\\
      &b&v&\\
      &v&b&\\
      &&&d
      \end{array}
   \right).
\end{equation}
Quantum correlations, as known (see, e.g, Ref.~\cite{MBCPV12}), must be invariant
under any local unitary transformations.
Take the local unitary transformation
\begin{equation}
   \label{eq:u}
   u=e^{i\pi\sigma_z/4}\otimes e^{-i\pi\sigma_z/4}
	 =\left(
      \begin{array}{cccc}
      1&&&\\
      &i&&\\
      &&-i&\\
      &&&1
      \end{array}
   \right).
\end{equation}
It is easy to check that the action of this transformation on the density matrix
(\ref{eq:rhoAB})
changes the sign of the off-diagonal entries, i.e., the matrix $u\rho_{AB}u^+$ is
equal to the right side of Eq.~(\ref{eq:rhoAB}) in which $v\to -v$.
Thus the density matrix can be brought to the form with real non-negative values
for all entries (more details for the general X density matrices can be found,
for example, in Refs.~\cite{Y14,Y14a}).
As a result, the matrix elements of density matrix (\ref{eq:rhoAB}) can be
written as
\begin{eqnarray}
   \label{eq:abdv}
   a=\frac{1}{Z}\exp[(J_z/2+B)/T],\quad
   b=\frac{1}{Z}e^{-J_z/2T}\cosh(J/T),
	 \nonumber\\
	 \\
   d=\frac{1}{Z}\exp[(J_z/2-B)/T],\quad
   v=\frac{1}{Z}e^{-J_z/2T}\sinh(|J|/T).
	 \nonumber
\end{eqnarray}
So, all properties of the system under question do not depend on the sign of
constant $J$, i.e., both ferro- and antiferro-interactions in transverse directions
of spin space lead to the same results.

Eigenvalues of density matrix $\rho_{AB}$ are equal to
\begin{equation}
   \label{eq:lam0}
   \lambda_1=a,\quad \lambda_2=d,\quad
   \lambda_{3,4}=b\pm v
\end{equation}
or, in an explicit form,
\begin{equation}
   \label{eq:lam}
   \lambda_{1,2}=\frac{1}{Z}\exp[(J_z/2\pm B)/T],\qquad
   \lambda_{3,4}=\frac{1}{Z}\exp[-(J_z/2\pm|J|)/T].
\end{equation}
The von Neumann entropy before measurement is
\begin{equation}
   \label{eq:S}
	 S(\rho_{AB})\equiv-{\rm Tr}\rho_{AB}\ln\rho_{AB}=-a\ln a -d\ln d
	 -(b+v)\ln(b+v)-(b-v)\ln(b-v).
\end{equation}
After inserting expressions for the quantities $a$, $b$, $d$, and $v$ this relation
returns to Eq.~(\ref{eq:S_TB}).

%----------------------------------------------------------------------
\subsection{
Post-measurement entropy and one-way deficit
}
\label{subsect:tildeS}
We now pursue a consideration of one-way quantum work deficit which was begun in
Introduction.
The maximum amount of useful work that can be extracted from the heat bath via working
body (substance) in the state $\rho$ is given as
\cite{OHHH02,HHHHOSS03,HHHOSSS05} (see also \cite{MBCPV12,Str15,ABC16})
\begin{equation}
   \label{eq:w}
   w=T(\log d_H - S(\rho)),
\end{equation}
where $S(\rho)=-{\rm tr}(\rho\log\rho)$ is the entropy of state $\rho$ and $d_H$ the
dimension of Hilbert space in which the density operator $\rho$ acts.
Applying this general relation to the states before and after Bob's measurement one
comes to the following working equation for the one-way quantum work (information)
deficit \cite{HHHOSSS05} (see also, e.g., \cite{MBCPV12,CCR15})
\begin{equation}
   \label{eq:tildeD}
   {\rm\Delta}=\min_{\{\rm\Pi_k\}}S(\tilde\rho_{AB})-S(\rho_{AB}),
\end{equation}
where
\begin{equation}
   \label{eq:rho_tilde}
   \tilde\rho_{AB}\equiv\sum_kp_k\rho^k_{AB}
	 =\sum_k({\rm I}\otimes{\rm\Pi}_k)\rho_{AB}({\rm I}\otimes{\rm\Pi}_k)^+
\end{equation}
is the weighted average of post-measured states
\begin{equation}
   \label{eq:rho-k}
   \rho^k_{AB}
	 =\frac{1}{p_k}({\rm I}\otimes{\rm\Pi}_k)\rho_{AB}({\rm I}\otimes{\rm\Pi}_k)^+
\end{equation}
with the probabilities
\begin{equation}
   \label{eq:p_k}
	 p_k={\rm Tr}({\rm I}\otimes{\rm\Pi}_k)\rho_{AB}({\rm I}\otimes{\rm\Pi}_k)^+.
\end{equation}
Thus, the one-way quantum work deficit equals the minimal increase of entropy after
nonselective orthogonal projective measurements on one party of the bipartite system.
Notice that the projective measurements increase entropy (but generalized measurements
can decrease it); the proof is based on Klein's inequality \cite{K31,NC10}.

So, like the quantum discord, the one-way deficit is non-negative quantity.
For the two-qubit systems, the entropies can vary from zero to two bits while the
one-way deficit may change its value from zero to one bit only.

As a matter of fact, the pre-measurement entropy $S$ in the expression for the one-way
deficit, Eq.~(\ref{eq:tildeD}), plays a roll of trivial shift.
Therefore, below we will stress attention mainly on the post-measurement entropies:
non-optimized $\tilde S(\theta)=S(\tilde\rho_{AB})$ and optimized
$\tilde{\cal S}=\min_\theta S(\tilde\rho_{AB})$ ones.

Further, operators ${\rm\Pi}_k$ in Eq.~(\ref{eq:rho_tilde}) are two projectors
($k=0,1$)
\begin{equation}
   \label{eq:Pi}
   {\rm\Pi}_k=V\pi_kV^+,
\end{equation}
where $\pi_k=|k\rangle\langle k|$ [i.e., $\pi_{0,1}=(1\pm\sigma_z)/2$] and
transformations $\{V\}$ belong to the special unitary group $SU_2$.
Rotations $V$ are parametrized by two angles $\theta$ and $\phi$ (polar and azimuthal,
respectively):
\begin{equation}
   \label{eq:V}
   V
	 =\left(
      \begin{array}{cc}
      \cos(\theta/2)&e^{i\phi}\sin(\theta/2)\\
      \sin(\theta/2)&-e^{i\phi}\cos(\theta/2)
      \end{array}
   \right)
\end{equation}
with $0\le\theta\le\pi$ and $0\le\phi<2\pi$.

Solving eigenproblem for the density matrix ${\tilde\rho}_{AB}$ defined by 
Eqs.~(\ref{eq:rhoAB}), (\ref{eq:abdv}), (\ref{eq:rho_tilde}), and (\ref{eq:V})
we get its eigenvalues
\begin{eqnarray}
   \label{eq:Lam1-4}
	 \Lambda_{1,2}=\frac{1}{4}\lbrack\!\lbrack1+(a-d)\cos\theta\pm\{[a-d+(1-4b)\cos\theta]^2
	 +4v^2\sin^2\theta\}^{1/2}\rbrack\!\rbrack
	 \nonumber\\
	 \\
	 \Lambda_{3,4}=\frac{1}{4}\lbrack\!\lbrack1-(a-d)\cos\theta\pm\{[a-d-(1-4b)\cos\theta]^2
	 +4v^2\sin^2\theta\}^{1/2}\rbrack\!\rbrack.
	 \nonumber
\end{eqnarray}
It is seen that the azimuthal angle $\phi$ has dropped out from the given expressions.
This is due to the fact that one pair of anti-diagonal entries of the density matrix
(\ref{eq:rhoAB}) vanishes.
Moreover, it is clear that every $\Lambda_i$ is invariant under the transformation
$v\to-v$ without replacement $v$ by $|v|$.
Using Eqs.~(\ref{eq:Lam1-4}) we arrive at the post-measured entropy (entropy after
measurement)
\begin{equation}
   \label{eq:postS}
   \tilde S(\theta;T, B)\equiv S(\tilde\rho_{AB})=h_4(\Lambda_1,\Lambda_2,\Lambda_3,\Lambda_4),
\end{equation}
where $h_4(x_1,x_2,x_3,x_4)=-\sum_{i=1}^4x_i\ln x_i$, with the additional condition
$x_1+x_2+x_3+x_4=1$, is the quaternary entropy function.
Notice that function $\tilde S$ of argument $\theta$ is invariant under the
transformation $\theta\to\pi-\theta$ therefore it is enough to consider the values for
which $\theta\in[0,\pi/2]$.

Completing this subsection we return to the second key property of non-optimized
one-way deficit which was announced in Introduction.
Simple calculations show that the first derivatives of post-measurement entropy,
Eqs.~(\ref{eq:Lam1-4}) and (\ref{eq:postS}), with respect to the angle $\theta$
identically equal zero at both endpoints $\theta=0$ and $\pi/2$ for any choice of $a$,
$b$, $d$, and $v$:
\begin{equation}
   \label{eq:postSD1}
   {\tilde S}^{\prime}(0)\equiv0,\quad
   {\tilde S}^{\prime}(\pi/2)\equiv0
\end{equation}
and consequently $\Delta^{\prime}(0)\equiv0$ and $\Delta^{\prime}(\pi/2)\equiv0$.

%----------------------------------------------------------------------
\subsection{
Piecewise-analytical-numerical formula for the one-way quantum work deficit
}
\label{sec:tildeD}
%......................................................................
\subsubsection{
Branch $\tilde S_0$
}
\label{sec:tildeS0}
Using Eqs.~(\ref{eq:Lam1-4}) and (\ref{eq:postS}) we find the post-measurement
entropy function by $\theta=0$:
\begin{equation}
   \label{eq:S0}
   \tilde S_0(T,B)\equiv \tilde S(0;T, B)=
	 -a\ln a-d\ln d-2b\ln b.
\end{equation}
This is a branch of entropy with the stationary optimal measurement angle equaled zero.

The entropy (\ref{eq:S0}) coincides the entropy of diagonal part of the density matrix
(\ref{eq:rhoAB}),
i.e., entropy of the Gibbs density matrix for the spin system (\ref{eq:H}) with
switched-off (fully suppressed) transverse interactions,
$\tilde\rho_{AB}|_{J=0}={\rm diag}[a,b,b,d]$.
This state corresponds to a purely classical, Ising system.
In other words, the nonselective $\sigma_z$-measurement
$\{{\rm\Pi}_k=\pi_k\ |\ k=0,1\}$
erases coherence (all non-diagonal terms) from the original density matrix.
After such a measurement, the entropy increases: $\tilde S_0(T,B)\ge S(T,B)$.
This is in agreement with a relation $S(\rho_{diag})\ge S(\rho)$
($\rho_{diag}$ is the diagonal part of arbitrary density matrix $\rho$) that follows
from the Klein inequality \cite{NC10} (see also \cite{F72}).

Notice that the entropy (\ref{eq:S0}) as well as the von Neumann entropy $S(T,B)$,
Eqs.~(\ref{eq:S_TB}) and (\ref{eq:S}), have the Gibbs forms and therefore they can be
measured by traditional way.
Namely, the specific heat $c(T)$ is first measured and then the ratio $c(T)/T$ is
integrated over the temperature: $S=\int_0^TdT[c(T)/T]$.

%......................................................................
\subsubsection{
Branch $\tilde S_{\pi/2}$
}
\label{sec:tildeS1}
The post-measured entropy function at the second endpoint $\theta=\pi/2$ equals
\begin{eqnarray}
   \label{eq:S1}
   \tilde S_{\pi/2}(T,B)\equiv \tilde S(\pi/2;T, B)
	 &&=-\frac{1+r}{2}\ln\frac{1+r}{4}-\frac{1-r}{2}\ln\frac{1-r}{4}
	 \nonumber\\
	 &&=\ln2+h((1+r)/2),
\end{eqnarray}
where
\begin{equation}
   \label{eq:r}
   r=[(a-d)^2+4v^2]^{1/2}
\end{equation}
and $h(x)=-x\ln x-(1-x)\ln(1-x)$ is the Shannon binary entropy function.

Here the nonselective measurement consists of projectors
${\rm\Pi}_{0,1}=(1\pm\sigma_x)/2$.
The state after this measurement equals (up to an insignificant angle $\phi$)
\begin{equation}
   \label{eq:rho_pi2}
   \tilde\rho_{AB}|_{\theta=\pi/2}
	 =\frac{1}{2}\!
	 \left(
      \begin{array}{cccc}
      a+b&.&.&v\\
      .&a+b&v&.\\
      .&v&b+d&.\\
      v&.&.&b+d
      \end{array}
   \right).
\end{equation}
The given matrix has the X form and its eigenvalues equal $\xi_{1,2}=(1\pm r)/4$,
each of them is two-fold degenerate.
Calculation of the von Neumann entropy returns us to Eq.~(\ref{eq:S1}).

According to the quantum mechanics,
``any result of a measurement of a real dynamical variable is one of its eigenvalues''
\cite{D58}.
On the other hand,
``the question now presents itself - Can every observable be measured?
The answer theoretically is yes.
In practice it may be very awkward, or perhaps even beyond the ingenuity of the
experimenter, to devise an apparatus which could measure some particular observable,
but the theory always allows one to imagine that the measurement can be
made'' \cite{D58}.
Unfortunately, so far no one has been able to make such a device to
measure the quantum von Neumann entropy, i.e., to project the operator $-\ln\rho$
(or, the same, $\rho$) into its diagonal representation.
In default of anything better, experimenters have developed a ``digital'' approach.
Namely, they perform quantum tomography of the state, numerically diagonalize it
(i.e, ``project'' into its eigenbasis), and then calculate the von Neumann entropy via
the eigenvalues found \cite{JKMW01}.

%......................................................................
\subsubsection{
Branch $\tilde S_{\vartheta}$
}
\label{sec:tildeS3}
The third possible branch is 
\begin{equation}
   \label{eq:S_vartheta}
   \tilde S_\vartheta(T,B)\equiv\tilde S(\vartheta,T,B)=\min_\theta\tilde S(\theta,T,B),
\end{equation}
where $\vartheta$ is the angle at which the post-measurement entropy reaches the
global minimum in the open interval $(0,\pi/2)$.
This branch is characterized by the variable, state-dependent optimal measurement
angle $\vartheta$ the value of which should be found numerically in practice.

So, the optimized post-measured entropy and one-way quantum work deficit are
piecewise-analytical-numerical functions which can be written as
\begin{equation}
   \label{eq:S_min}
   \tilde{\cal S}(T,B)=\min\{\tilde S_0,\tilde S_\vartheta,\tilde S_{\pi/2}\}
\end{equation}
and
\begin{equation}
   \label{eq:rmDelta}
   {\rm\Delta}(T,B)=\min\{\Delta_0,\Delta_\vartheta,\Delta_{\pi/2}\},
\end{equation}
respectively.
Here the branches $\tilde S_0$ and $\tilde S_{\pi/2}$ are known in analytical forms
and only the third branch $\tilde S_\vartheta$ requires to solve one-dimensional
optimization problem.

%----------------------------------------------------------------------
\subsection{
Equations for the possible boundaries 
}
\label{sect:EqBound}
The question now arises: how to find boundaries between different branches of one-way
quantum work deficit?
The answer to this question is reduced to the following.

If in some region of domain of the one-way deficit function there are only the
branches $\tilde S_0$ and $\tilde S_{\pi/2}$ (global interior minimum is absent) then
the equation for the boundary between them is obviously defined as
\begin{equation}
   \label{eq:S0S1}
   \tilde S_0=\tilde S_{\pi/2}.
\end{equation}
When crossing the boundary, the optimal measurement angle suddenly changes its value
by $\pi/2$ while the function itself stays continuous, but it can have a fracture on
its curve \cite{Y19a}.
Taking into account the analytical expressions for these branches, Eqs.~(\ref{eq:S0})
and (\ref{eq:S1}), we come to a transcendental equation which can be solved
numerically, for example, by the bisection method.

Further, if the branch with the interior global minima is present then the situation
is more complicated.
The first attempt to locate the region with such a branch for the quantum discord
was undertaken in Ref.~\cite{CZYYO11}.
The authors of this paper found several sufficient conditions for the branches with
the optimal observables $\sigma_z$ and $\sigma_x$ and established that the
intermediate region (with the interior optimal measurement angles) can separate the
regions with the former two branches.
However, nothing was said about the boundaries between different regions and from
their figures it was not clear whether such boundaries are precise or, v.v., smoothed
(as, for example, in a rainbow between colors).
The same concerns the later paper \cite{H13}.

Both quantum discord $\cal Q$ and one-way deficit $\rm\Delta$, Eq.~(\ref{eq:rmDelta}),
are piecewise-defined functions and the boundaries between their branches (phases) are
precise.
Equations for the boundaries with the intermediate regions were first proposed in
Refs.~\cite{Y14,Y14a} (again  for the quantum discord; an extension to the one-way
deficit is obvious).
The idea was based on the next observation.
If the function of $\theta$ (it doesn't matter the non-optimized discord or one-way
deficit) is {\em unimodal} (i.e., it may have only one local extremum) in the open
interval $(0,\pi/2)$ then by varying parameters the local extremum can come inside
such an interval or goes out only trough the endpoints $\theta=0$ or $\pi/2$
\cite{Y14,Y14a} (see also \cite{Y15}).
This is clear from the view point of topology (``rubber geometry'').
At a result, the equations for the boundaries based on the unimodality hypothesis are
given as
\begin{equation}
   \label{eq:Sii0_0}
   \tilde S^{\prime\prime}(0)=0
\end{equation}
(for the 0-boundary) and
\begin{equation}
   \label{eq:Sii1_0}
   \tilde S^{\prime\prime}(\pi/2)=0
\end{equation}
(for the $\pi/2$-boundary).
When crossing these boundaries, the optimal measurement angle stays invariable,
its jump $\Delta\vartheta=0$.

At the 0- and $\pi/2$-boundaries, the second derivatives of the
post-measurement entropy (\ref{eq:postS}) with respect to the angle $\theta$ are given
by
\begin{equation}
   \label{eq:Sii0}
   \tilde S^{\prime\prime}_0=\tilde S^{\prime\prime}(0)=\frac{1}{4}\left[(a-d)\ln\frac{a}{d}+(1-4b)\ln\frac{ad}{b^2}
	 -2v^2\left(\frac{1}{a-b}\ln\frac{a}{b}+\frac{1}{b-d}\ln\frac{b}{d}\right)\right]
\end{equation}
and
\begin{eqnarray}
   \label{eq:Sii1}
   \tilde S^{\prime\prime}_{\pi/2}=\tilde S^{\prime\prime}(\pi/2)&&=8v^2\frac{ab+bd-ad+v^2}{r^3}\ln\frac{1+r}{1-r}
	 \nonumber\\
	 &&-\frac{1}{2}(a-d)^2\left\{\frac{[1+(1-4b)/r]^2}{1+r}+\frac{[1-(1-4b)/r]^2}{1-r}\right\}\!\!,\
\end{eqnarray}
where $r$ is given, as before, by Eq.~(\ref{eq:r}).
Equating these expressions to zero we obtain transcendental equations for the
boundaries under discussion.

In Ref.~\cite{Y18} (see also \cite{Y19}), it has been discovered that the one-way
deficit $\Delta(\theta)$ may also have in some regions a {\em bimodal} behavior, i.e.,
it exhibits two extrema (one minimum and one maximum) in the open interval
$(0,\pi/2)$.
This leads to the additional boundary, $0^\prime$, which obeys the equation
\begin{equation}
   \label{eq:SvarS0}
   \tilde S_0=\tilde S_\vartheta.
\end{equation}
Inserting into this equality the corresponding expressions for the branches, we arrive
at the transcendental equation which now requires at every step of bisection process
to search the optimal measurement angle $\vartheta$;
the latter can be made by using, for example, the golden section method.
At the boundary the optimal measurement angle $\vartheta$ jumps a finite step
$\Delta\vartheta<\pi/2$.
Solution of Eq.~(\ref{eq:SvarS0}) reproduces the solution of Eq.~(\ref{eq:Sii0_0})
when the jump $\Delta\vartheta$ tends to zero.

It is interesting to note that the continue and discontinue behavior of the optimal
measurement angle $\vartheta$ at the boundaries is similar to the behavior of order
parameter in the Landau theory of phase transitions.
These parallels as well as a relation with the bifurcation phenomena and catastrophe
theory will be discussed below, in Sec.~\ref{sect:bifurc}.

It can also be noted that, in principle, the equation
$\tilde S_\vartheta=\tilde S_{\pi/2}$ is possible,
but the corresponding boundary does not occur in our research.

%======================================================================
\section{
Phase diagrams and discussion
}
\label{sec:PhDs}
We now turn to finding and classifying all qualitatively different types
of temperature-field phase diagrams for the model (\ref{eq:H}) when changing the
parameter $J_z/|J|$ from minus infinity to plus infinity.

Our approach consists of two steps.
First, we solve the above boundary equations and draw the obtained lines on the plane
$(T,B)$.
Then, using the shape of curve $\tilde S(\theta)$, we identify the character of phase
in each subregion of $(T,B)$-plane.

%----------------------------------------------------------------------
\subsection{
The case $J_z=-|J|$
}
\label{sect:Jz-1}
\begin{figure}[t]
\begin{center}
\epsfig{file=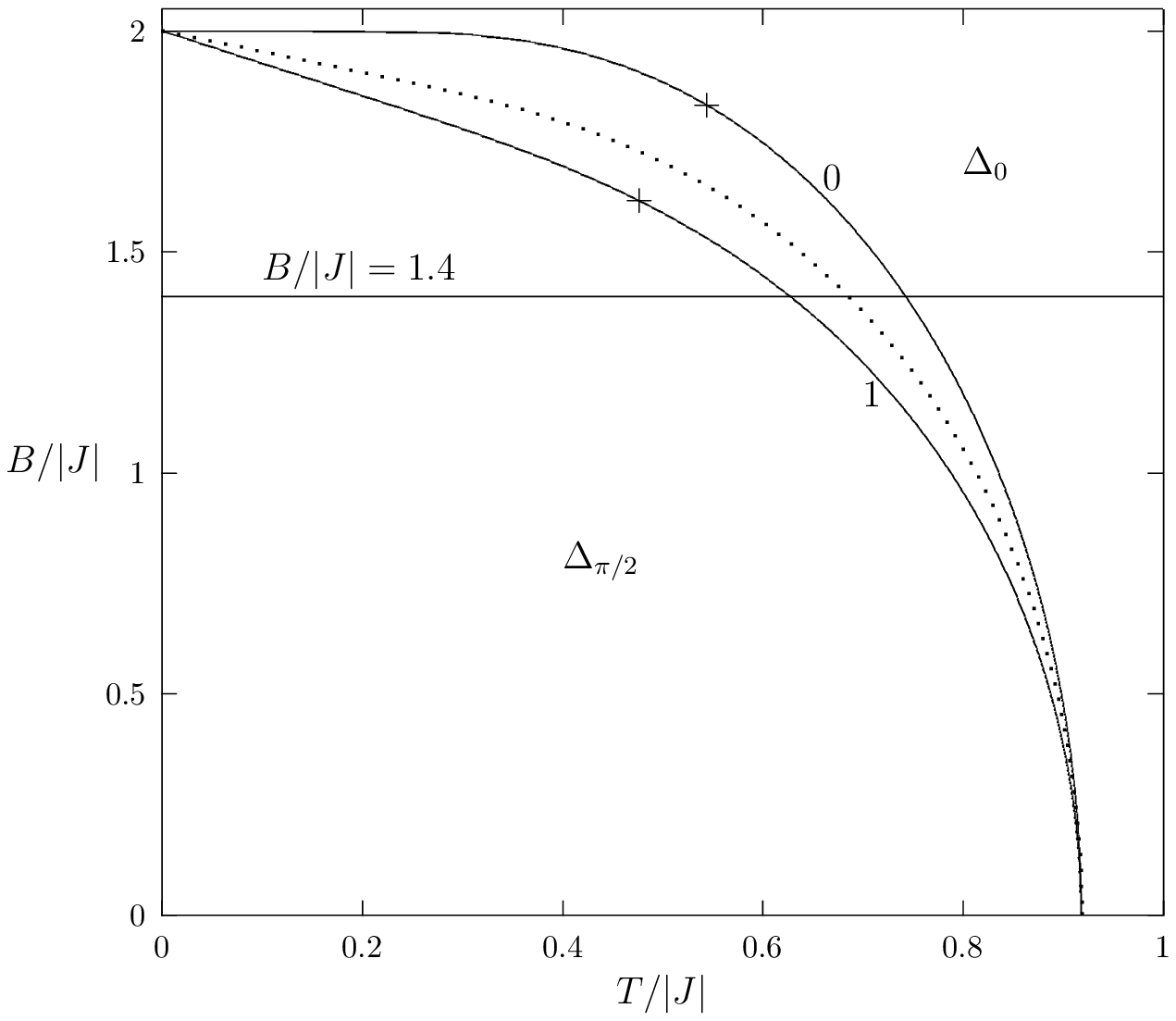,width=5.6cm}
\hspace{3mm}
\epsfig{file=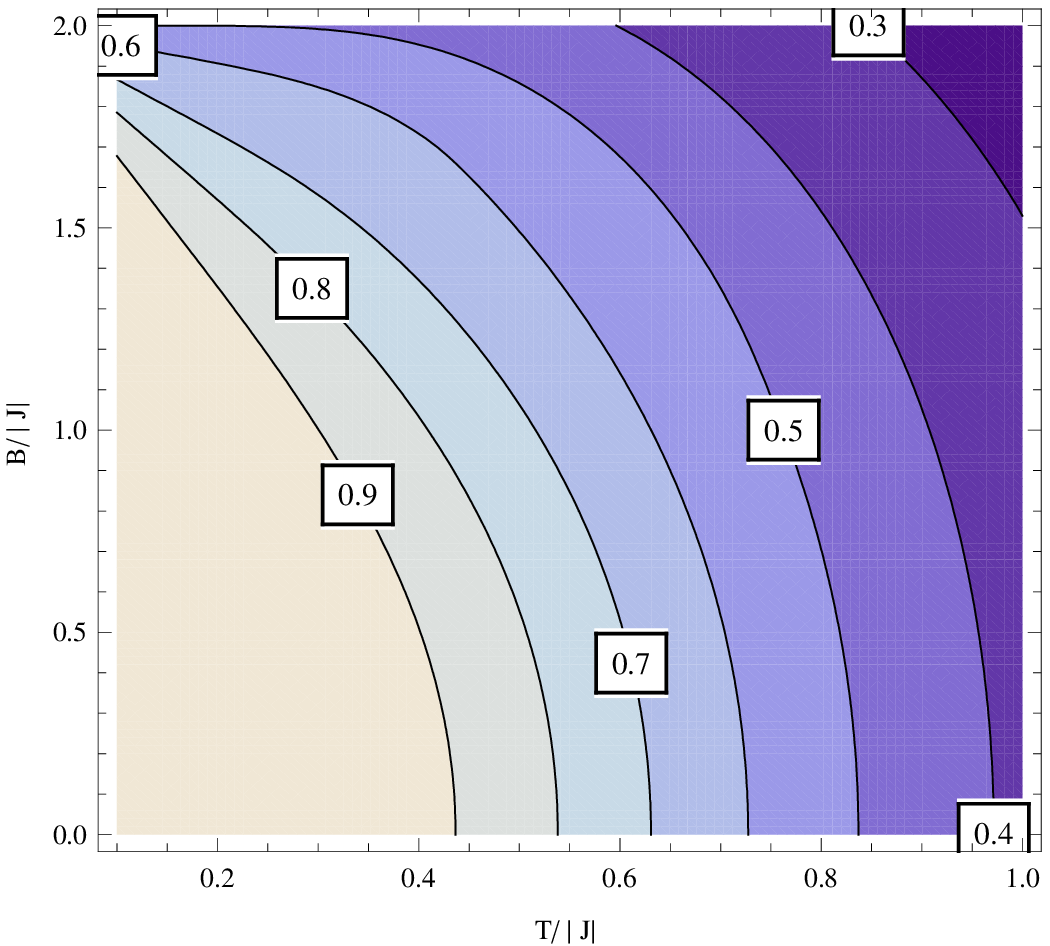,width=5.4cm}
\caption{
(Color online)
Temperature-field phase diagram (left) and level lines (right) for the one-way deficit
of the model under question with $J=\pm1$ and $J_z=-1$.
The fraction $\Delta_\vartheta$ is located between the solid lines 0 and 1 on the
phase diagram;
dotted line corresponds to the condition $\Delta_0=\Delta_{\pi/2}$ and does not here
serve as the boundary.
The strait line $B/|J|=1.4$ is a path for probing the shape of $\tilde S(\theta)$.
Two pluses ($+$) on the boundaries 0 an 1 mark the points for characterizing the size
of $\Delta_\vartheta$-region
}
\label{fig:z1m1a}
\end{center}
\end{figure}

Let us start with the antiferromagnetic coupling $J_z$ equal to $-|J|$.
Having solved Eqs.~(\ref{eq:S0S1})-(\ref{eq:Sii1_0}) and (\ref{eq:SvarS0}), we get the
graph shown on the left in Fig.~\ref{fig:z1m1a}.
%......................................................................
%                          FIGURE 1
%\begin{figure}[h]
%......................................................................
The lines 0 and 1 are the 0- and $\pi/2$-boundaries corresponding to the solutions of
Eqs.~(\ref{eq:Sii0_0}) and (\ref{eq:Sii1_0}), respectively.
The dotted line results from Eq.~(\ref{eq:S0S1}).
Solution of Eq.~(\ref{eq:SvarS0}) reproduces the solution of Eq.~(\ref{eq:Sii0_0}).
These three lines tend to $T/|J|=0.91758$ when $B/|J|\to0$ (see Fig.~\ref{fig:z1m1a},
on the left).

\begin{figure*}[t]
\begin{center}
\epsfig{file=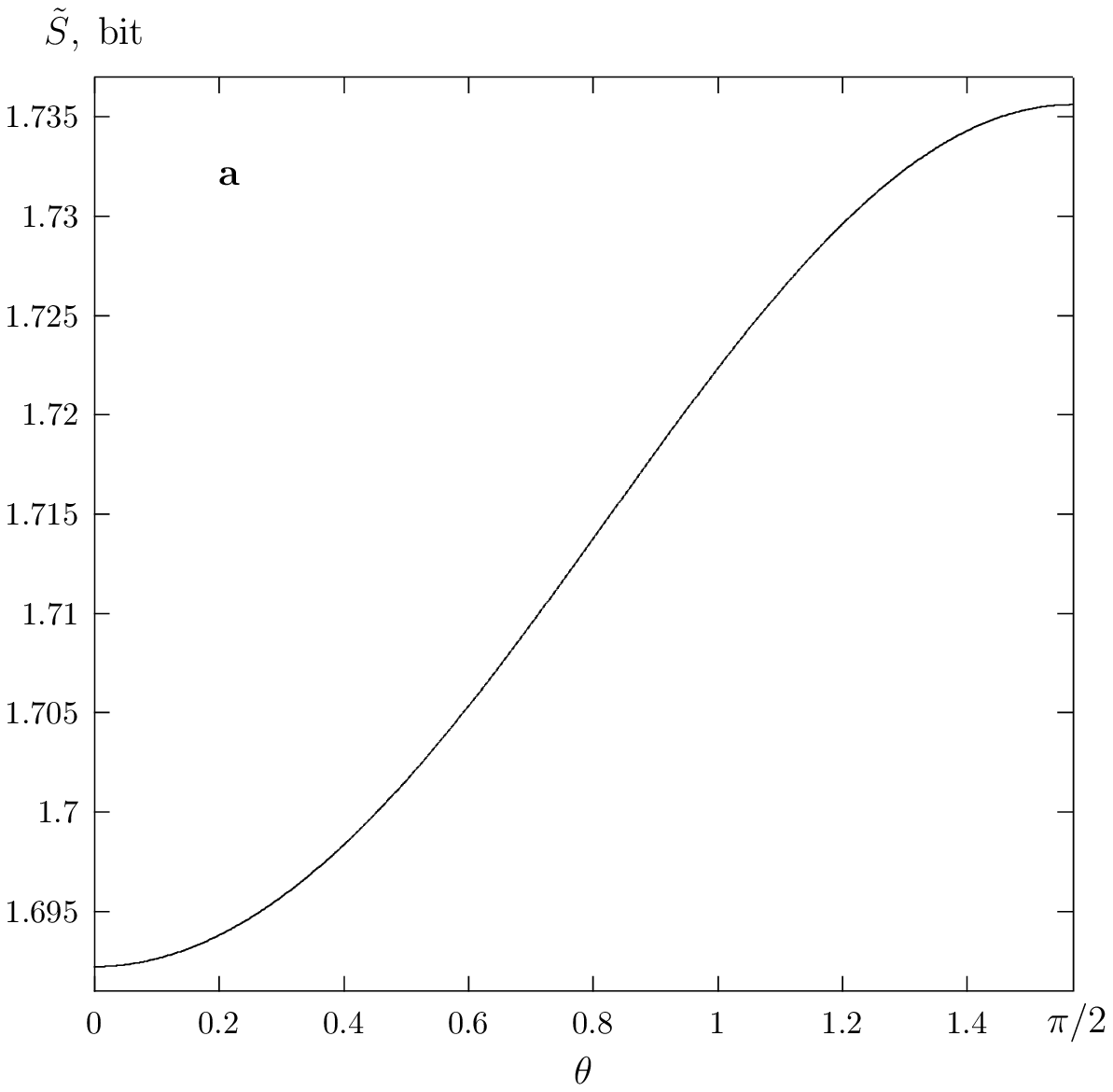,width=3.8cm}
\epsfig{file=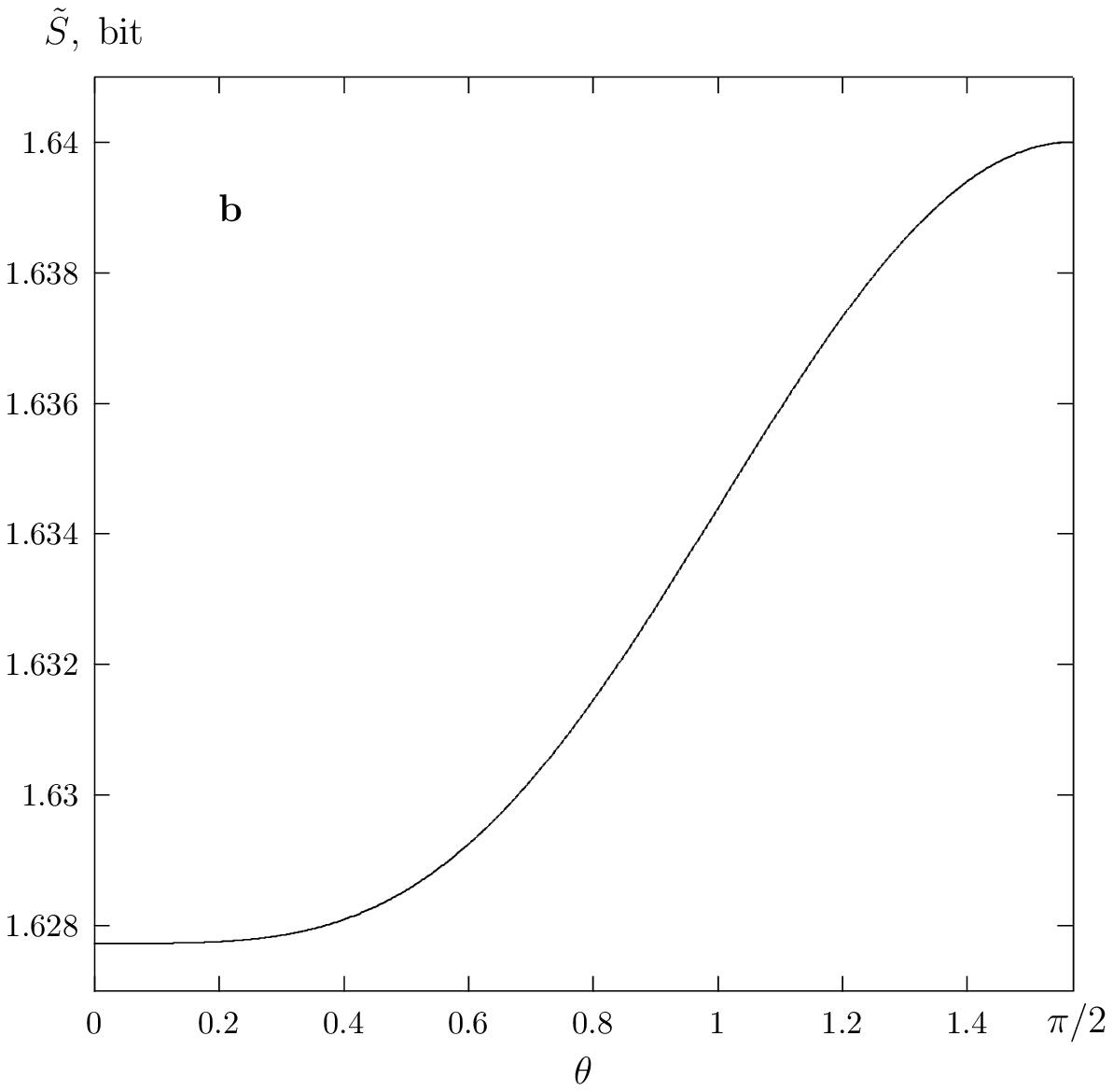,width=3.8cm}
\epsfig{file=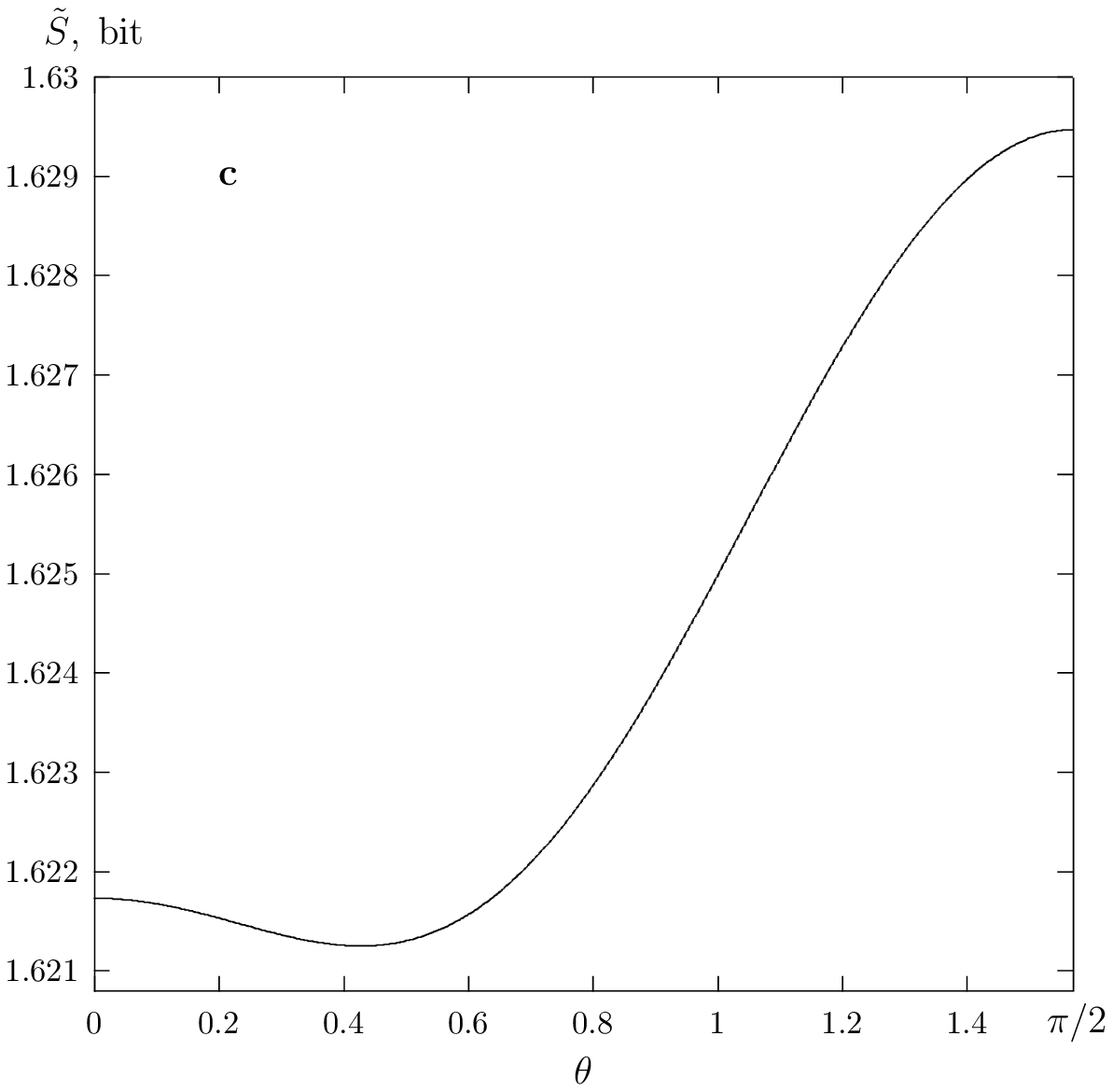,width=3.8cm}
\epsfig{file=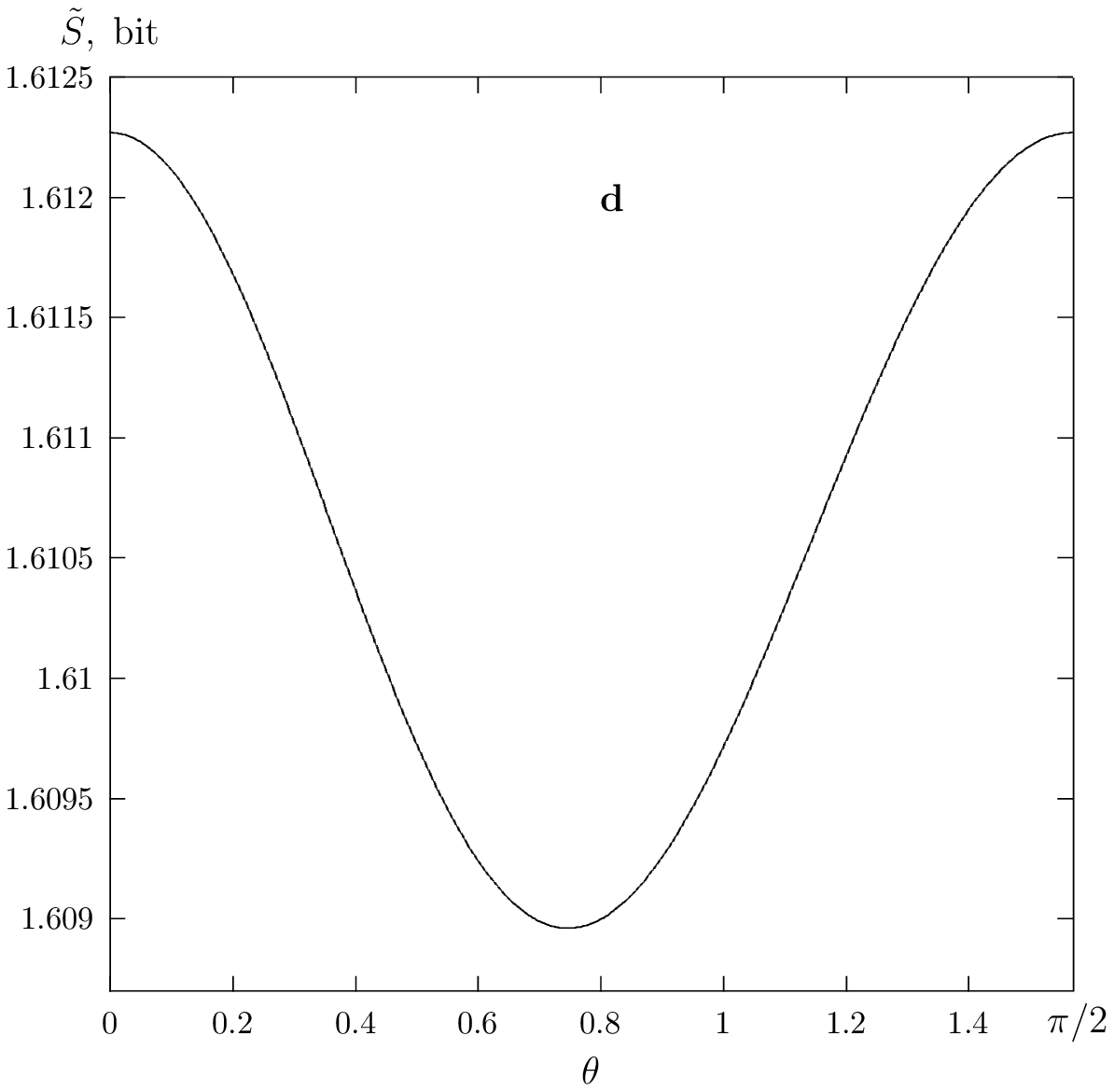,width=3.8cm}
\epsfig{file=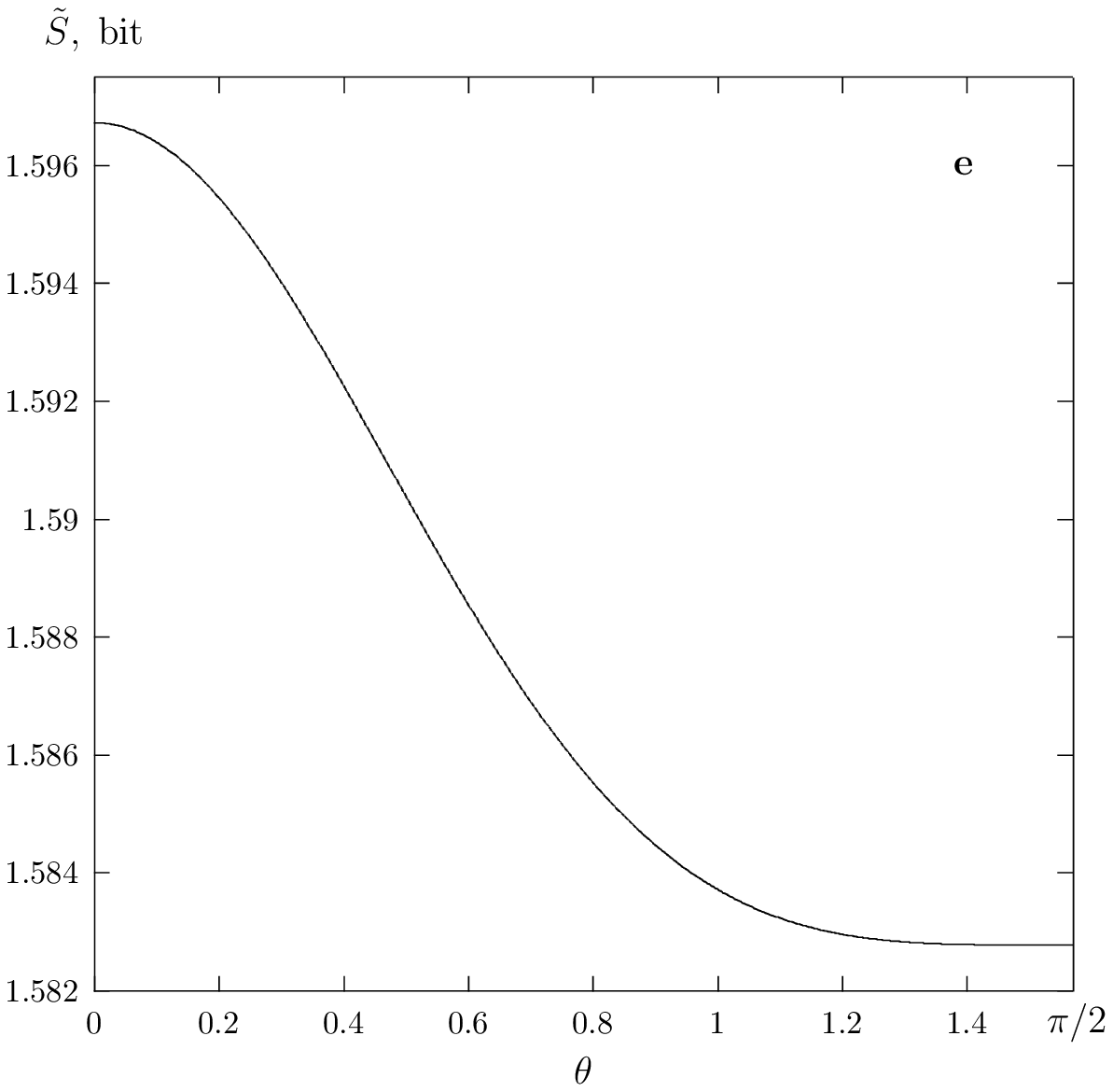,width=3.8cm}
\epsfig{file=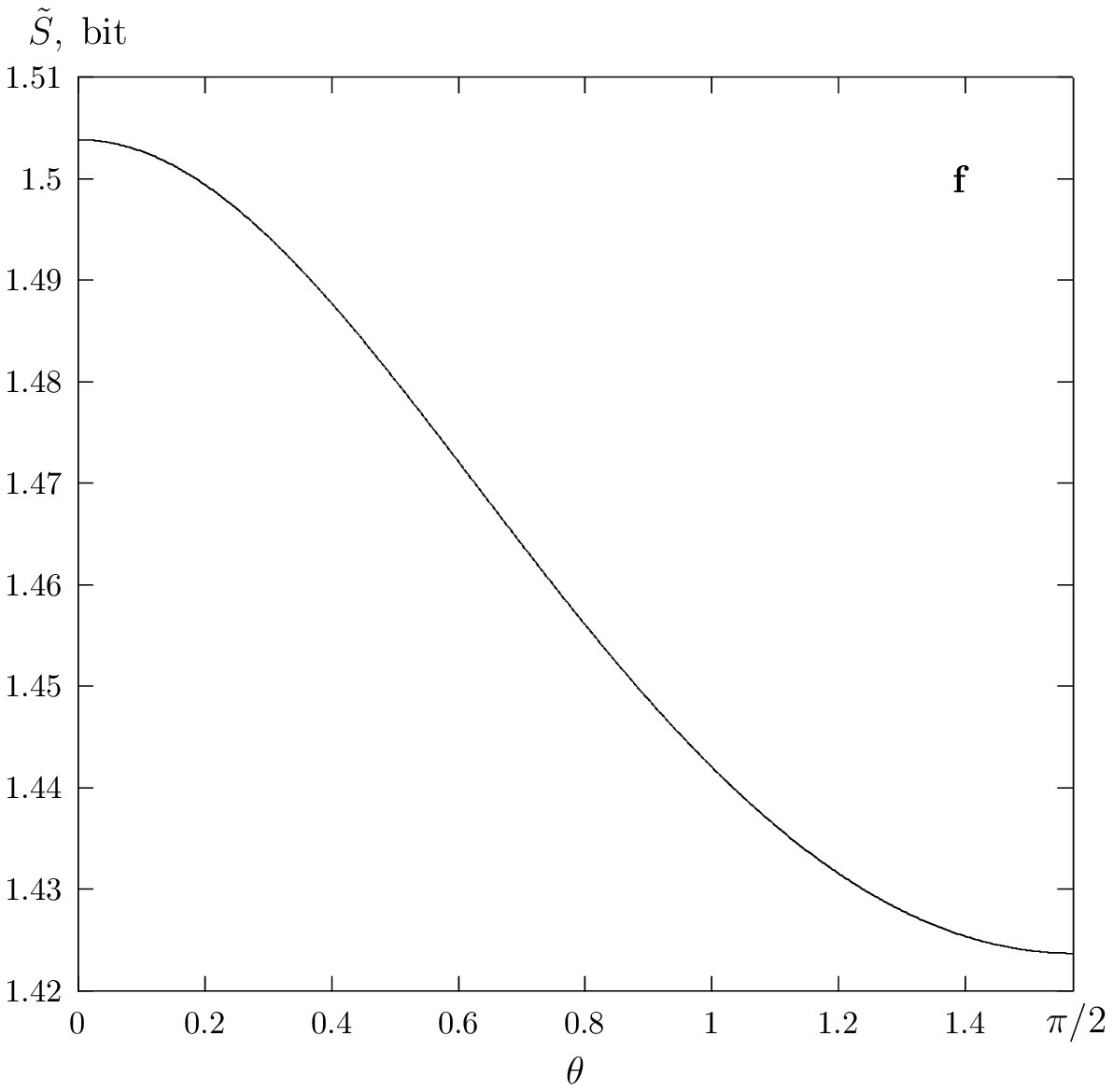,width=3.8cm}
\caption{
Post-measurement entropy $\tilde S$ vs $\theta$
by $J=-1$, $J_z=-1$, $B=1.4$
and $T=1$~(a), 0.742967~(b), 0.72~(c), 0.684237~(d), 0.6275~(e),
0.4~(f)
}
\label{fig:zs1mB14}       % Give a unique label
\end{center}
\end{figure*}

To identify different regions on the plane $(T,B)$ we will study the shapes of
post-measurement entropy function $\tilde S(\theta)$ in various points of the plane.
Let us take, for example, a straight line $B/|J|=1.4$ (see Fig.~\ref{fig:z1m1a},
left).
Evolution of $\tilde S(\theta)$ by moving along this path is shown in
Fig.~\ref{fig:zs1mB14}.
%
%......................................................................
%                          FIGURE 2
%......................................................................

When the argument $T$ is large enough the curve $\tilde S(\theta)$ exhibits a
monotonically increasing shape as shown in Fig.~\ref{fig:zs1mB14}~(a).
Here a position of minimum lies at $\theta=0$ and does not depend on $T$
(stationary optimal measurement angle).
This means that the region $\Delta_0$ is here.
With decreasing $T$ the minimum of the post-measurement entropy at the point
$\theta=0$ becomes broader, see Fig.~\ref{fig:zs1mB14}~(b).
Further decreasing the temperature bears an interior minimum
[see Fig.~\ref{fig:zs1mB14}~(c)] position of which depends now on $T$ what tells us
that we have entered into the region $\Delta_\vartheta$ with the variable optimal
measurement angle $\vartheta$.
At the temperature $T=0.684237$ the path reaches the dotted line, where 
$\tilde S_0=\tilde S_{\pi/2}$ as clear seen in Fig.~\ref{fig:zs1mB14}~(d).
Then the interior minimum merges with the minimum at the endpoint $\theta=\pi/2$;
this situation is shown in Fig.~\ref{fig:zs1mB14}~(e).
After this the region $\Delta_{\pi/2}$ with the constant optimal measurement angle
$\theta=\pi/2$ takes place up to $T=0$.
So, we have classified all subregions and hence obtain the phase diagram by
$J_z/|J|=-1$.
 
It is interesting to estimate the characteristic width of the region with the
variable optimal measurement angle.
For this purpose we chose two points on the boundaries 0 and $\pi/2$
marked be symbols ``plus'' $(+)$ that have coordinates $(0.5444,1.8323)$
and $(0.4765,1.6164)$, see the left Figure~\ref{fig:z1m1a}.
To find the width, one may use the fidelity $F$ which is related to the Bures distance
$d_B$ by the equation $d_B=[2(1-\sqrt{F})]^{1/2}$.
Since the density matrices $\rho_{AB}$ by different values of $a$, $b$, $d$ and $v$
commute, the fidelity is given as
\begin{equation}
   \label{eq:fid}
   F=\biggl(\sum_{j=1}^4\sqrt{\lambda_j^{(1)}\lambda_j^{(2)}}\biggr)^{\!\!2}.
\end{equation}
Here $\lambda_j^{(1,2)}$ are the eigenvalues (\ref{eq:lam}) for the two
points taken on the plane $(T,B)$.
Using Eq.~(\ref{eq:fid}) we get $F=0.985645$ ($\approx98.6\%)$ between the above
points.
One may conclude from here that the region with the variable optimal measurement angle
is broad enough and can be detected on an experiment (see in this connection
\cite{MGY19}).

The graph on the right side of Fig.~\ref{fig:z1m1a} shows the distribution of one-way
quantum work deficit values together with the level lines.
The maximum value $\rm\Delta=1$ is achieved at the point $(0,0)$ and then, how it is
clear seen from this figure, the quantum correlation asymptotically tends to zero when
$T$ and $B$ increase.

%----------------------------------------------------------------------
\subsection{
Bifurcations and catastrophes at the boundaries
}
\label{sect:bifurc}
In this subsection, we will make a small digression and discuss in more detail the
mechanism of the appearance and disappearance of interior extrema in the function
$\tilde S(\theta)$.

\begin{figure}[t]
%\begin{figure}
\begin{center}
\epsfig{file=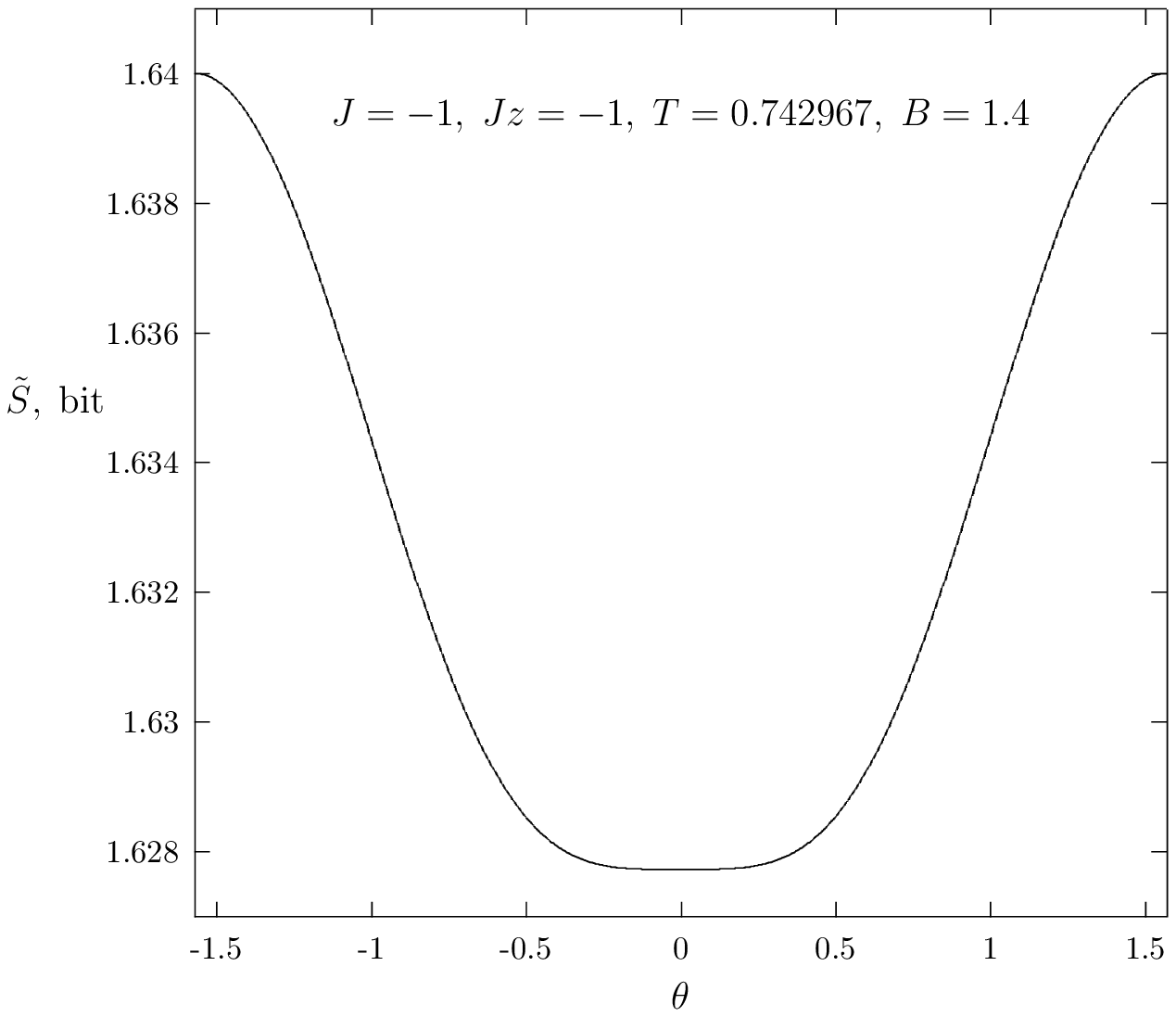,width=5.4cm}
\hspace{3mm}
\epsfig{file=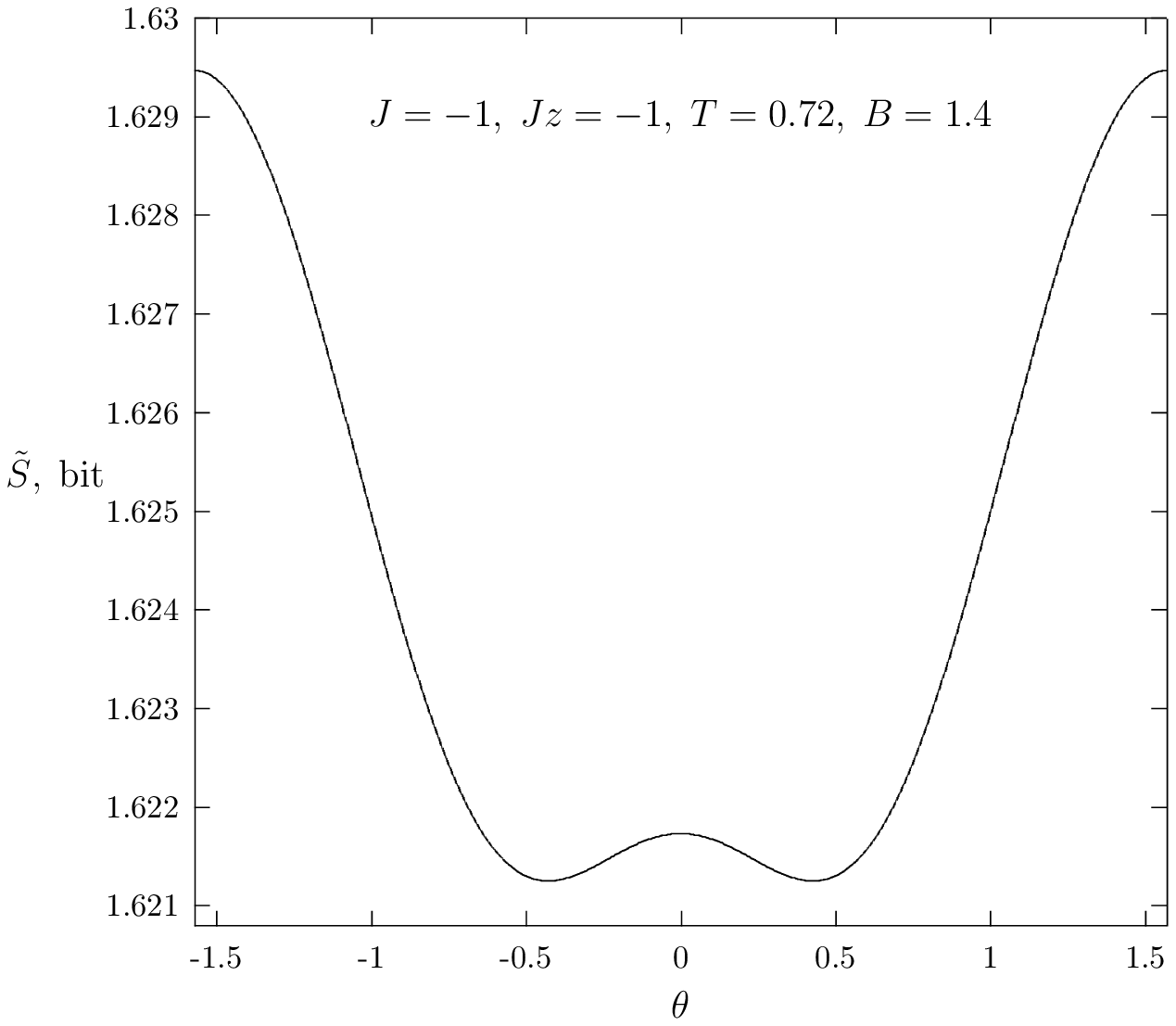,width=5.4cm}
\caption{
Post-measurement entropy $\tilde S$ vs $\theta$ in the window $[-\pi/2,\pi/2]$ for the
model by $J_z/|J|=-1$, $B/|J|=1.4$, and $T=T_c=0.742967$ (left) and $T=0.72$ (right).
The bifurcation of the minimum at $\theta=0$ is clear visible:
the behavior of function like $y\propto x^4$ (left) changes to
$y\propto x^4-\epsilon x^2$ (right)
}
\label{fig:zs1mT072B14a}
\end{center}
\end{figure}

To this end, let us look at the behavior of this function from the previous example
($J_z/|J|=-1$ and $B/|J|=1.4$) but in a wider window, namely
$\theta\in[-\pi/2,\pi/2]$.
Graphs of $\tilde S(\theta)$ at and near the 0-boundary are shown in
Fig.~\ref{fig:zs1mT072B14a}.
%......................................................................
%                          FIGURE 3
%\begin{figure}[h]
%......................................................................
When the temperature is large enough the optimal measurement angle is a constant equal
to zero and the minimized post-measurement entropy is described by the branch
$\tilde S_0(T)$.
Near the minimum $\theta=0$, the function $\tilde S(\theta)\sim\theta^2$.
As the temperature $T\to T_c=0.742967$, the minimum broadens and at the point
$T=T_c$ the function becomes like $\tilde S(\theta)\sim\theta^4$
(Fig.~\ref{fig:zs1mT072B14a}, left). 
However, with an arbitrary small further decrease in $T$ a ``catastrophe'' occurs:
the local minimum at $\theta=0$ suddenly splits (bifurcates) into two ones 
(see Fig.~\ref{fig:zs1mT072B14a}, right), and the minimized post-measurement entropy,
Eq.~(\ref{eq:S_min}),
changes (jumps) from $\tilde S_0(T)$ to the another branch $\tilde S_\vartheta(T)$.

The above picture is quantitatively described as follows.
The function $\tilde S(\theta)$ is even and its expansion in a Tailor series at
$\theta=0$ has the form
\begin{equation}
   \label{eq:S_thet}
   \tilde S(\theta; T,B)=\tilde S_0(T,B)
	 +\frac{1}{2!}\tilde S^{\prime\prime}_0(T,B)\!\cdot\!\theta^2
	 +\frac{1}{4!}\tilde S^{iv}_0(T,B)\!\cdot\!\theta^4+\ldots,
\end{equation}
where $\tilde S_0(T,B)$ and $\tilde S^{\prime\prime}_0(T,B)$ are given by
Eqs.~(\ref{eq:S0}) and (\ref{eq:Sii0}), respectively.
This series is similar to the Landau expansion of the thermodynamic potential
(Gibbs free energy) in his theory of second order phase transitions \cite{LL_StPh}.
In our problem, $\tilde S(\theta)$ plays a role of the thermodynamic potential and
the optimal measurement angle $\vartheta$ is an analogue of the order parameter in
 Landau's theory.
The vanishing of the coefficient at the quadratic term corresponds to the phase
transition point.

Various sudden changes in behavior of a system as the variables that control it are
changed continuously are studied and classified in the catastrophe theory \cite{A92}.
Regarding our problem, it is useful to analyze the behavior of the function 
\begin{equation}
   \label{eq:f-1}
	 f(\theta)=-t\theta^2+\theta^4.
\end{equation}
For $\theta>0$, this function is either monotonically increasing (when the parameter
$t\le0$) or unimodal (has one minimum), when $t>0$.
In the latter case, the minimum is located at $\vartheta=\sqrt{t/2}$ and its depth
equals $h=t^2/4$.
Hence, the distance to this minimum increases with the speed
$\dot\vartheta=1/(2\sqrt{2t})$, and the depth of the minimum grows with the speed
$\dot h=t/2$.
Thus, at the moment of birth $(t=0)$, the position of the minimum moves with infinite
speed, whereas the depth of the minimum increases extremely slowly - with zero speed.
The former is favorable to numerical search the interior minimum but the latter, on
the contrary, makes the task difficult.
Therefore, for finding the 0-boundary, it is better to use Eq.~(\ref{eq:Sii0_0}) than
Eq.~(\ref{eq:SvarS0}).
Notice, the expansion of the quantity $t(T)$ at the point $T=T_c$ up to the first
nonzero term is equal to $t=\alpha_0(T-T_c)$.

Return again to the discussion of example from the previous subsection.
When moving along the path $B/|J|=1.4$, the interior minimum reaches the
$\pi/2$-boundary at the temperature $T=0.6275$ (see Fig.~\ref{fig:z1m1a}).
Here the inner minimum merges with the minimum at the bound $\theta=\pi/2$ as shown in
Fig.~\ref{fig:zs1mB14}~(e).
With further decreasing the temperature, the optimal measurement angle stays $\pi/2$
up to zero absolute temperature.
So, one may conclude that, if the conditional or post-measurement entropy function is
unimodal, then its local extremum can appear inside the open interval $(0,\pi/2)$ or,
v.v., disappear in it only if the extremum enters or leaves trough one of two
boundaries: at 0 or $\pi/2$.
It was the observation that allowed in Refs.~\cite{Y14,Y14a,Y15} to propose the
boundary equations, based on the second derivatives, for the regions with the variable
optimal measurement angle.

Later, however, the bimodal behavior has been discovered in some cases for the one-way
quantum work deficit \cite{Y18}.
In the next subsection, we will meet such situations.
To describe mathematically the bimodal behavior one should to expand the function
under optimization up to sixth order terms:
\begin{equation}
   \label{eq:f-2}
   f(\theta)=\alpha_1\theta^2+\alpha_2\theta^4+\theta^6.
\end{equation}
For $\theta>0$, this function can have {\em two} local extrema: one minimum and one
maximum.
Under some conditions, the additional minimum can drop lower that the minimum at
$\theta=0$.
Then a new phenomenon arises, namely, the optimal measurement angle 
discontinuously changes/jumps from zero to the angle corresponding to the inner
minimum  (see \cite{Y18,Y19} and the next subsection).
In the context of Landau's theory, similar finite jumps of the order parameter
corresponds the first order phase transitions.

This concludes the digression and we move on to discussing the types of phase diagrams
again.

%----------------------------------------------------------------------
\subsection{
The case $J_z<-|J|$
}
\label{sect:Jz<-1}
\begin{figure}[t]
%\begin{figure}
\begin{center}
\epsfig{file=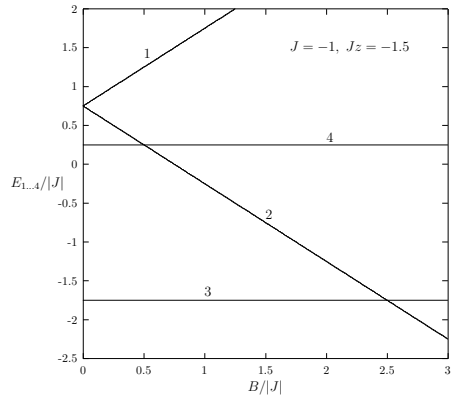,width=5.8cm}
\caption{
Splitting scheme of energy levels for the model with $J=-1$ and $J_z=-1.5$.
Lines $1-4$ correspond to the energies $E_1-E_4$, respectively
}
\label{fig:ze15m}
\end{center}
\end{figure}

Consider now the case when the coupling constant $J_z$ is negative and dominates in
magnitude.
This corresponds to the antiferromagnetic Ising-like systems.
Let, for definiteness, J be equal to $-1$ and $J_z=-1.5$.
In this case, the energy spectrum, Eq.~(\ref{eq:Ei}), consists of two singlets and one
doublet that is split into two levels by an external magnetic field as shown in
Fig.~\ref{fig:ze15m}.
%......................................................................
%                          FIGURE 4
%\begin{figure}[h]
%......................................................................
The energy level $E_2=-J_z/2-B$ crosses the singlet levels $E_3=J_z/2+J$ and
$E_4=J_z/2-J$ at $B/|J|=2.5$ and $0.5$, respectively.

\begin{figure}[t]
%\begin{figure}
\begin{center}
\epsfig{file=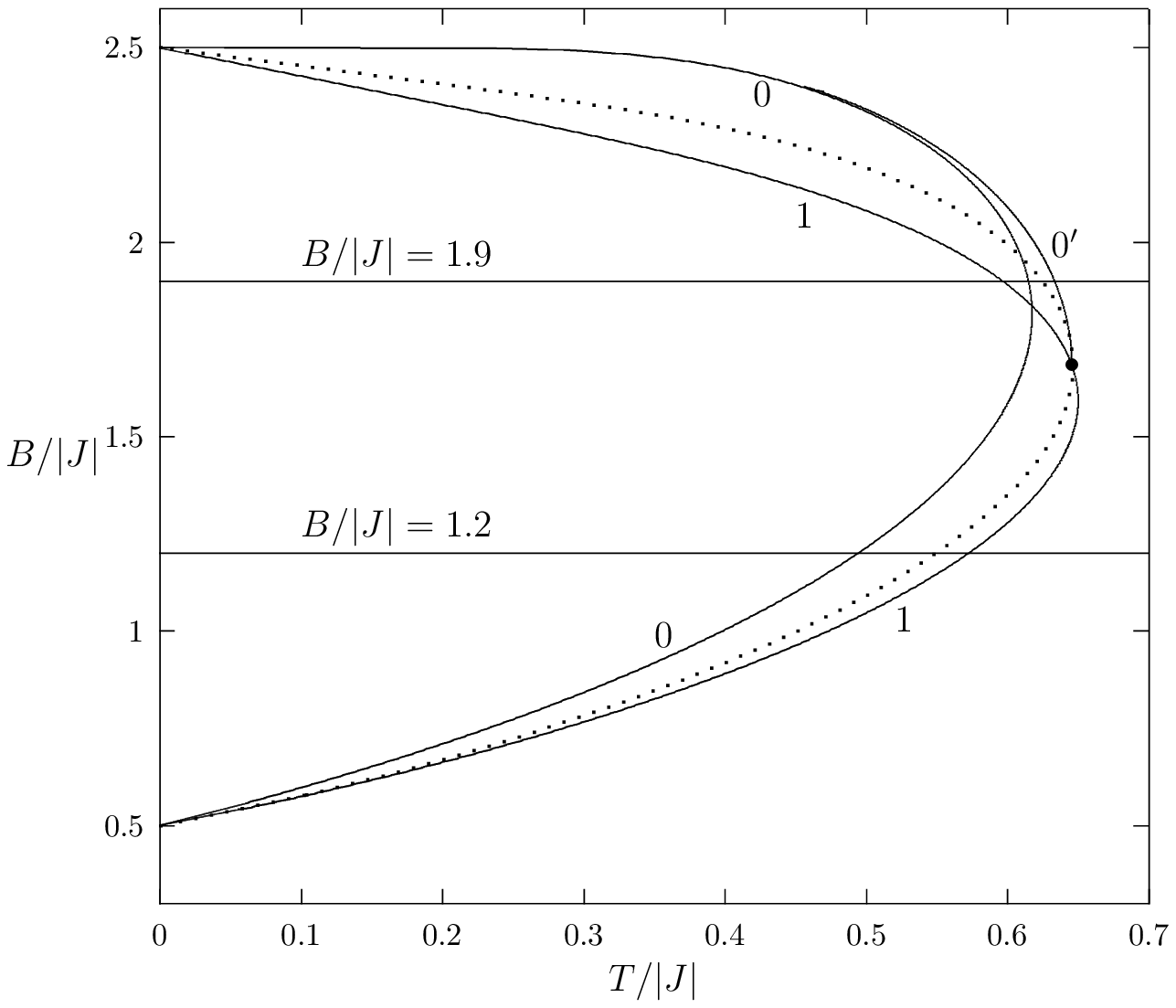,width=5.4cm}
\hspace{3mm}
\epsfig{file=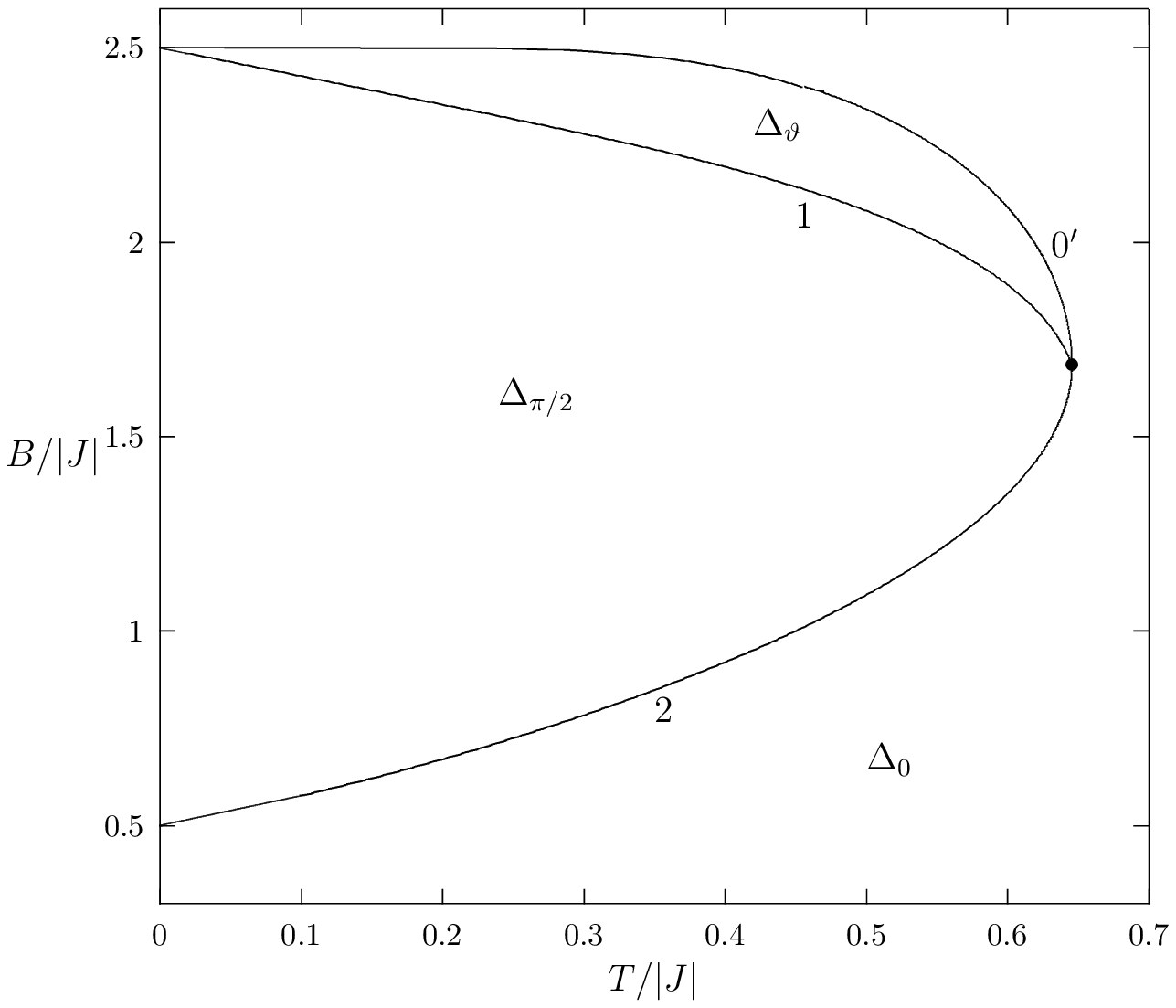,width=5.4cm}
\caption{
On the left, the curves corresponding to the solutions of all boundary equations and,
on the right, the phase diagram of the one-way deficit for the system with $|J|=1$ and
$J_z=-1.5$.
Black circle ($\bullet$) is the triple point.
Straight lines $B/|J|=1.9$ and $B/|J|=1.2$ are the paths along which the shapes of
$\tilde S(\theta)$ are studied (see Figs.~\ref{fig:zs15mB19} and \ref{fig:zs15mB12})
}
\label{fig:z15my}
\end{center}
\end{figure}

Formal solution of the boundary equations (\ref{eq:S0S1}) - (\ref{eq:Sii1_0}) and
(\ref{eq:SvarS0}) leads to the curves shown on the left in Fig.~\ref{fig:z15my} by the
dotted, 0, 1, and $0^\prime$ lines, respectively.
%
%......................................................................
%                          FIGURE 5
%\begin{figure}[h]
%......................................................................
%
The dotted curve, a candidate on the boundary $\Delta_0=\Delta_{\pi/2}$, and the
curves 1 and $0^\prime$ intersect at the point $(0.6454108,1.6851637)$ marked by the
black circle.

\begin{figure*}[t]
\begin{center}
%\vspace{1cm}
%\hspace{1cm}
\epsfig{file=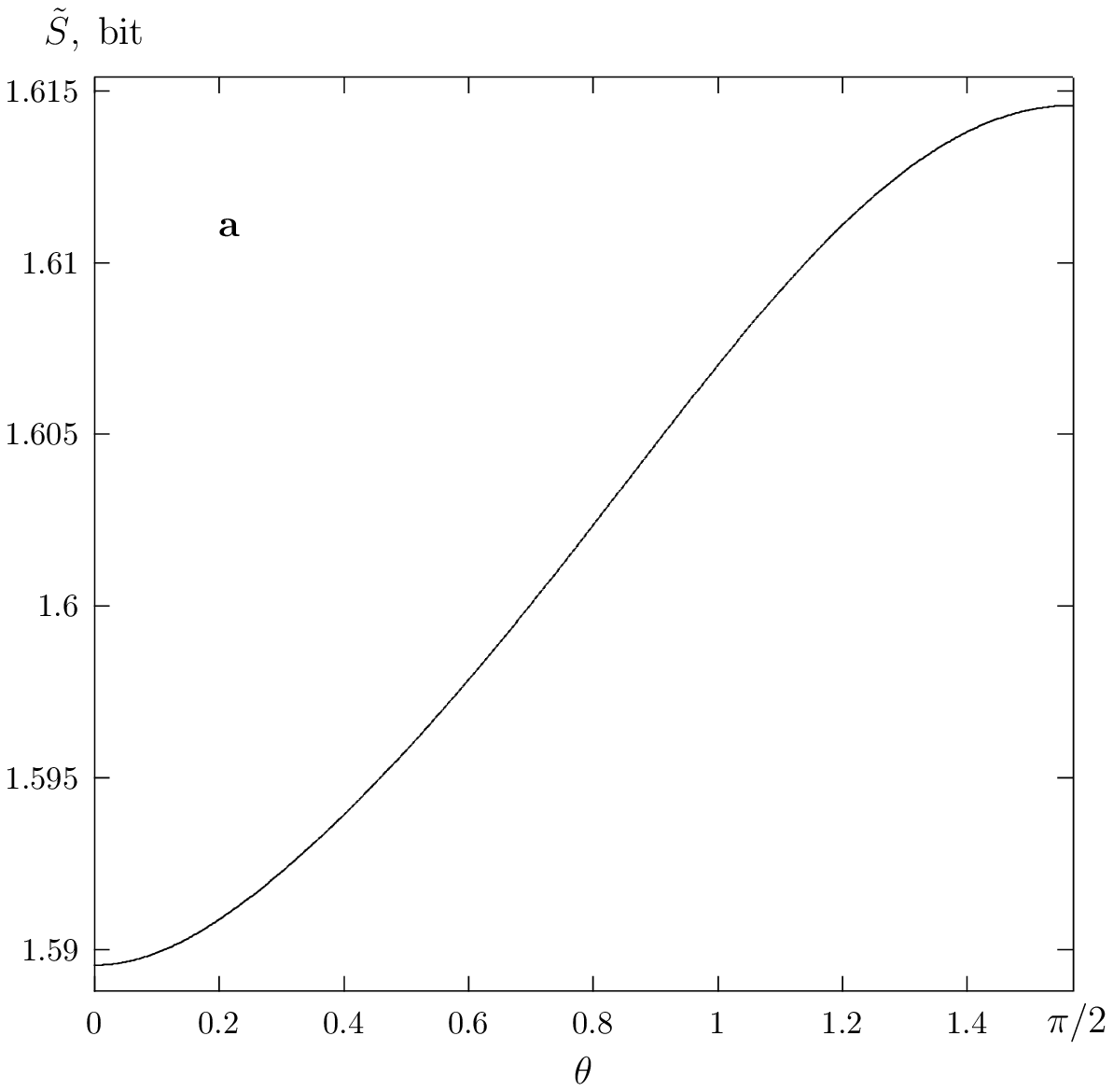,width=3.8cm}
\epsfig{file=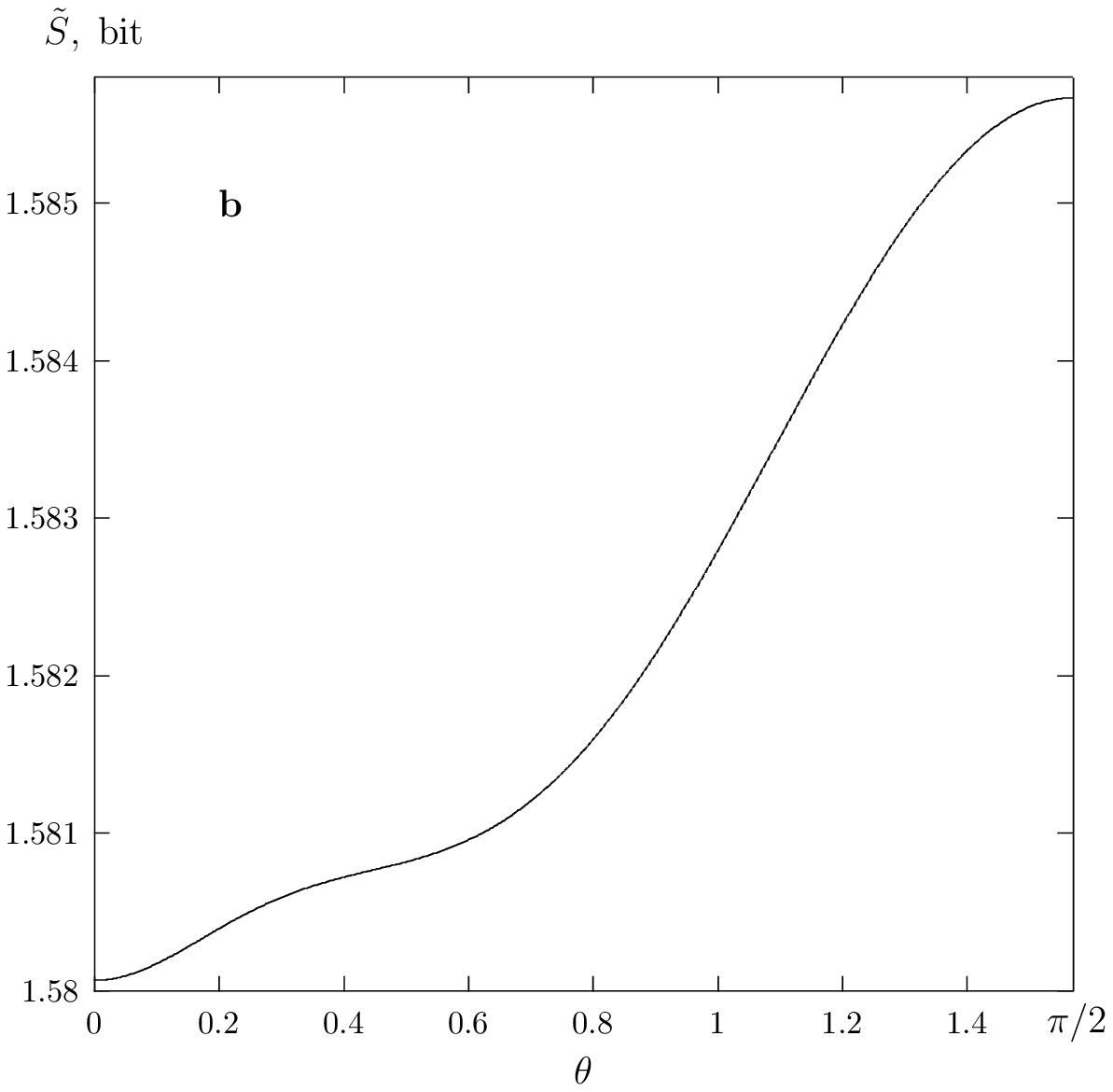,width=3.8cm}
\epsfig{file=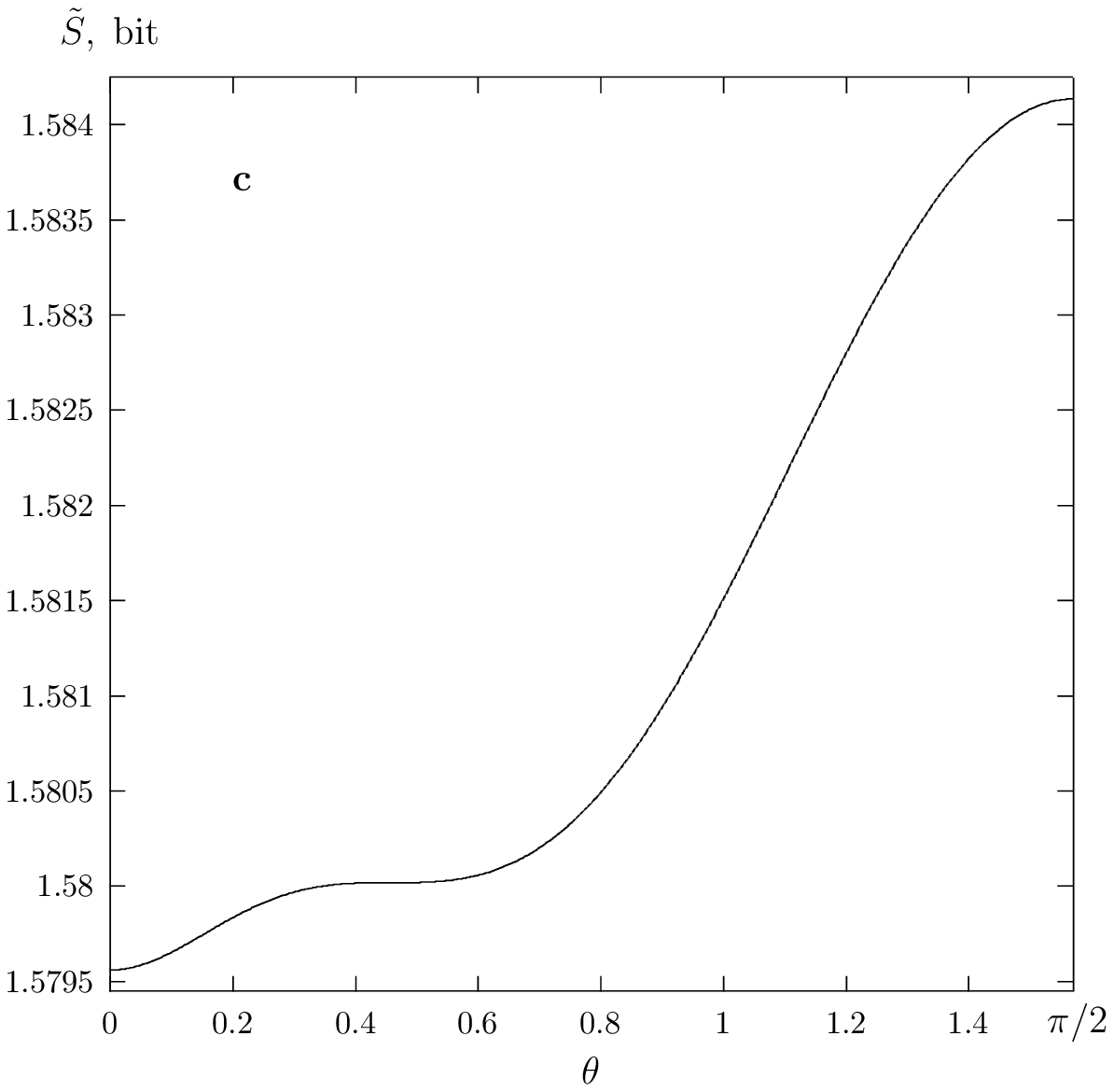,width=3.8cm}
\epsfig{file=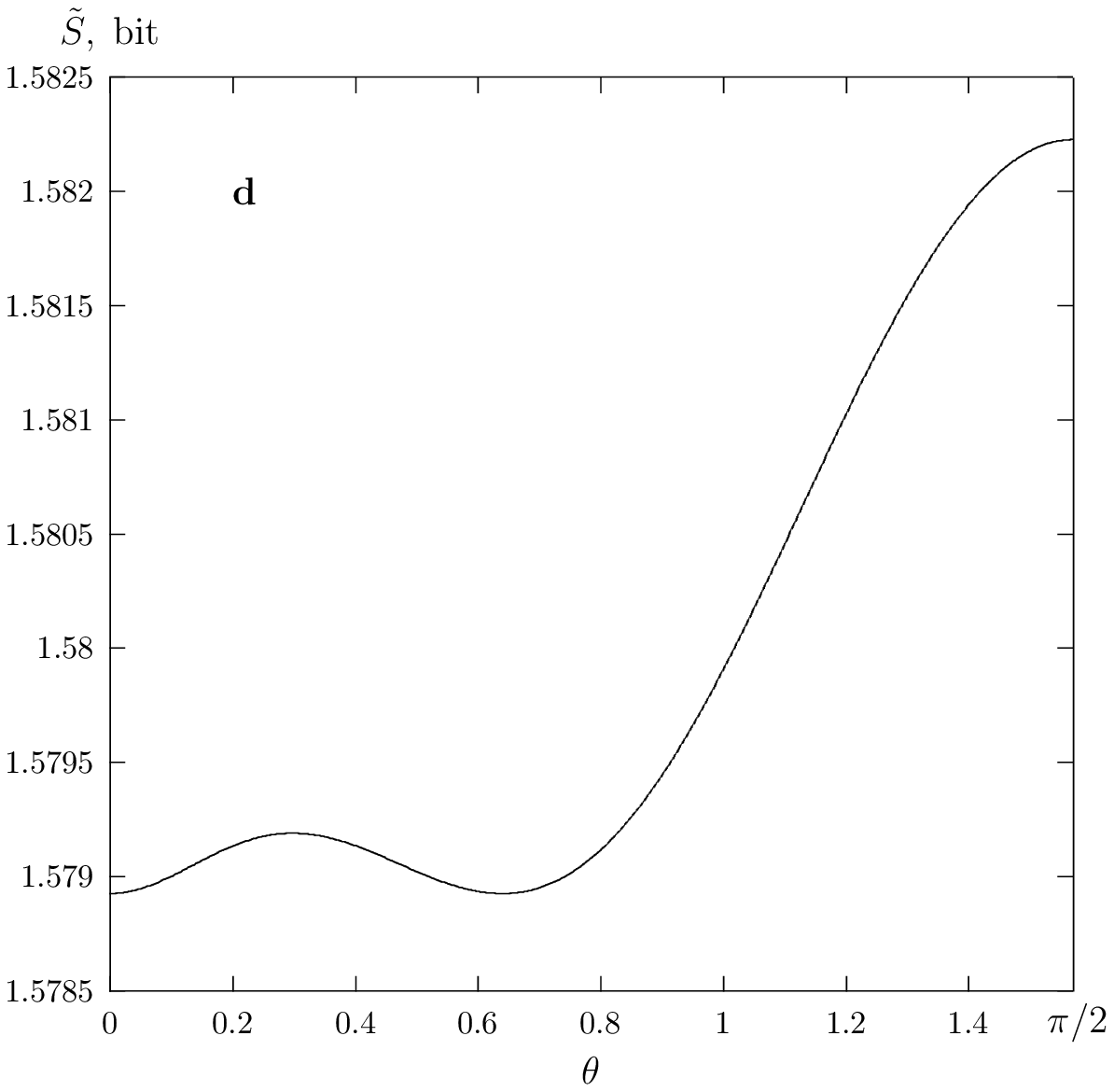,width=3.8cm}
\epsfig{file=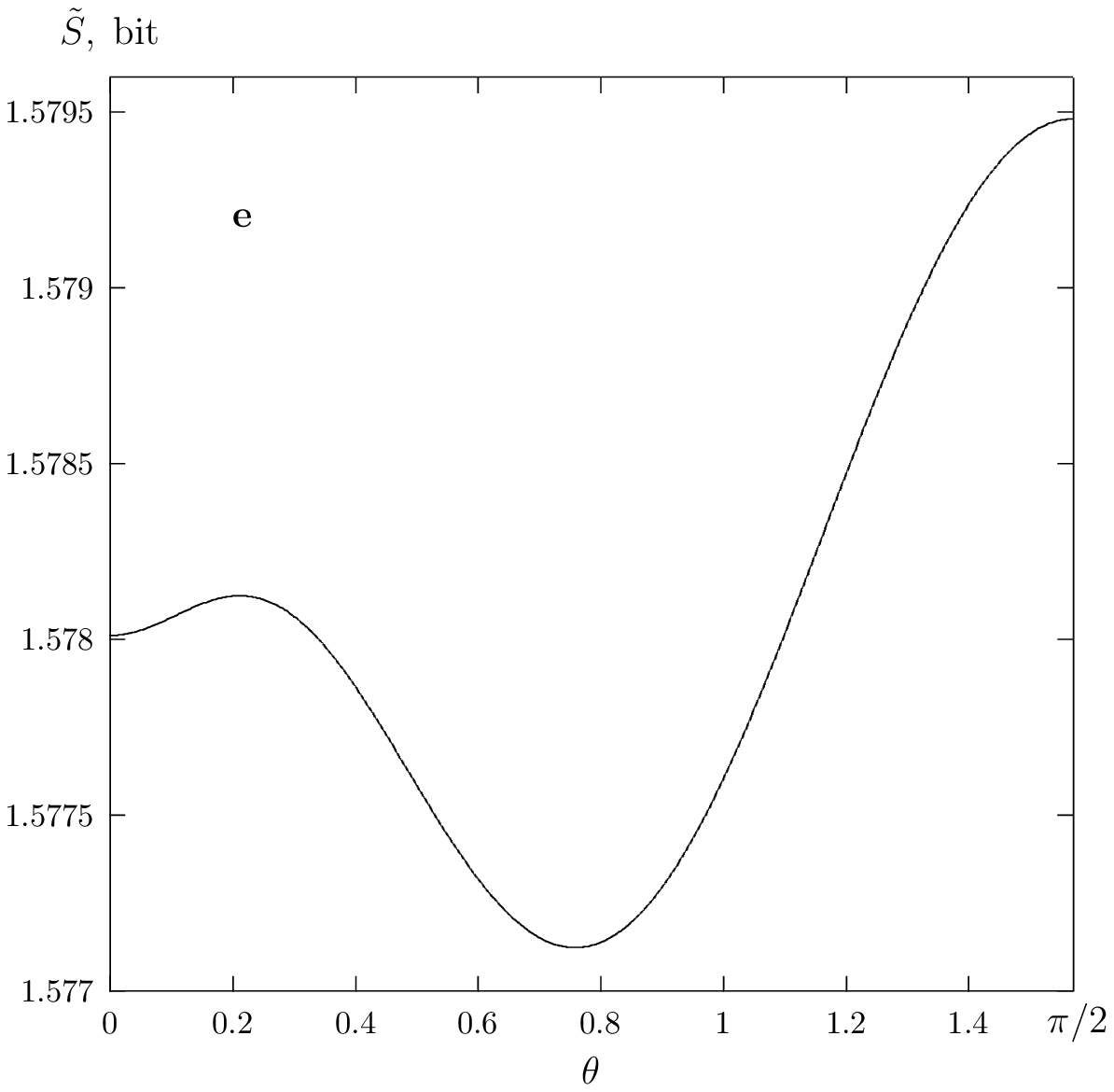,width=3.8cm}
\epsfig{file=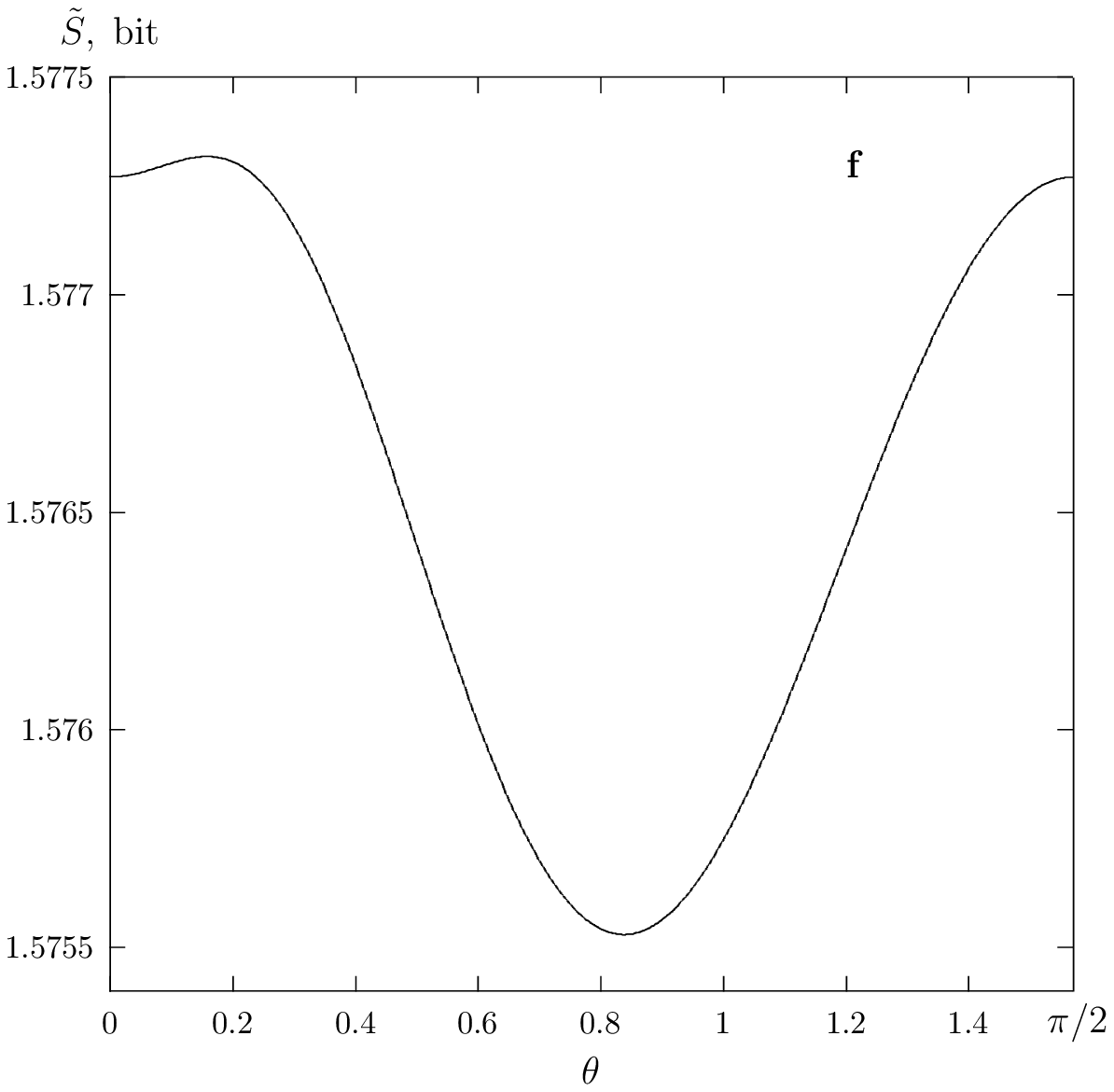,width=3.8cm}
\epsfig{file=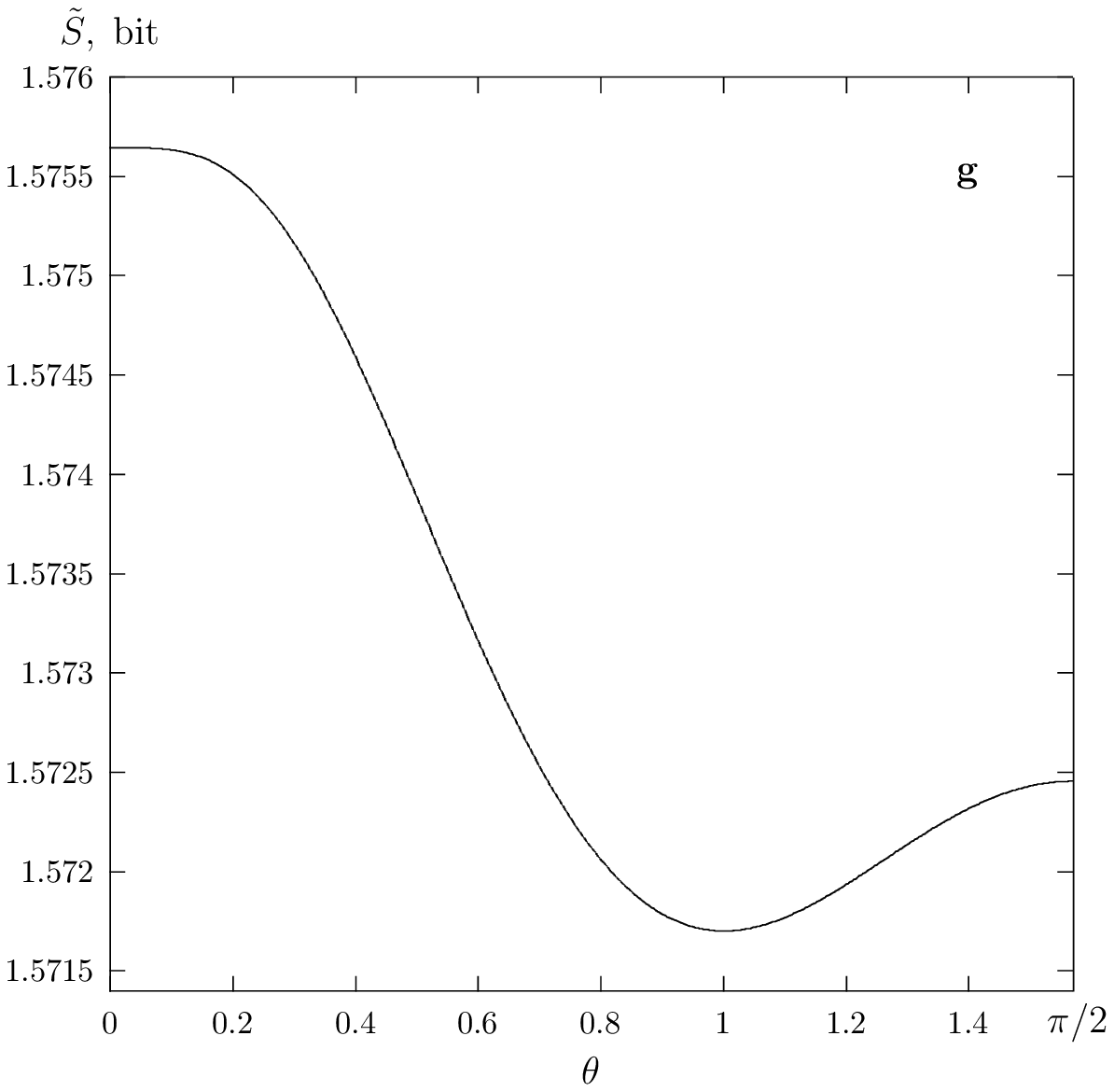,width=3.8cm}
\epsfig{file=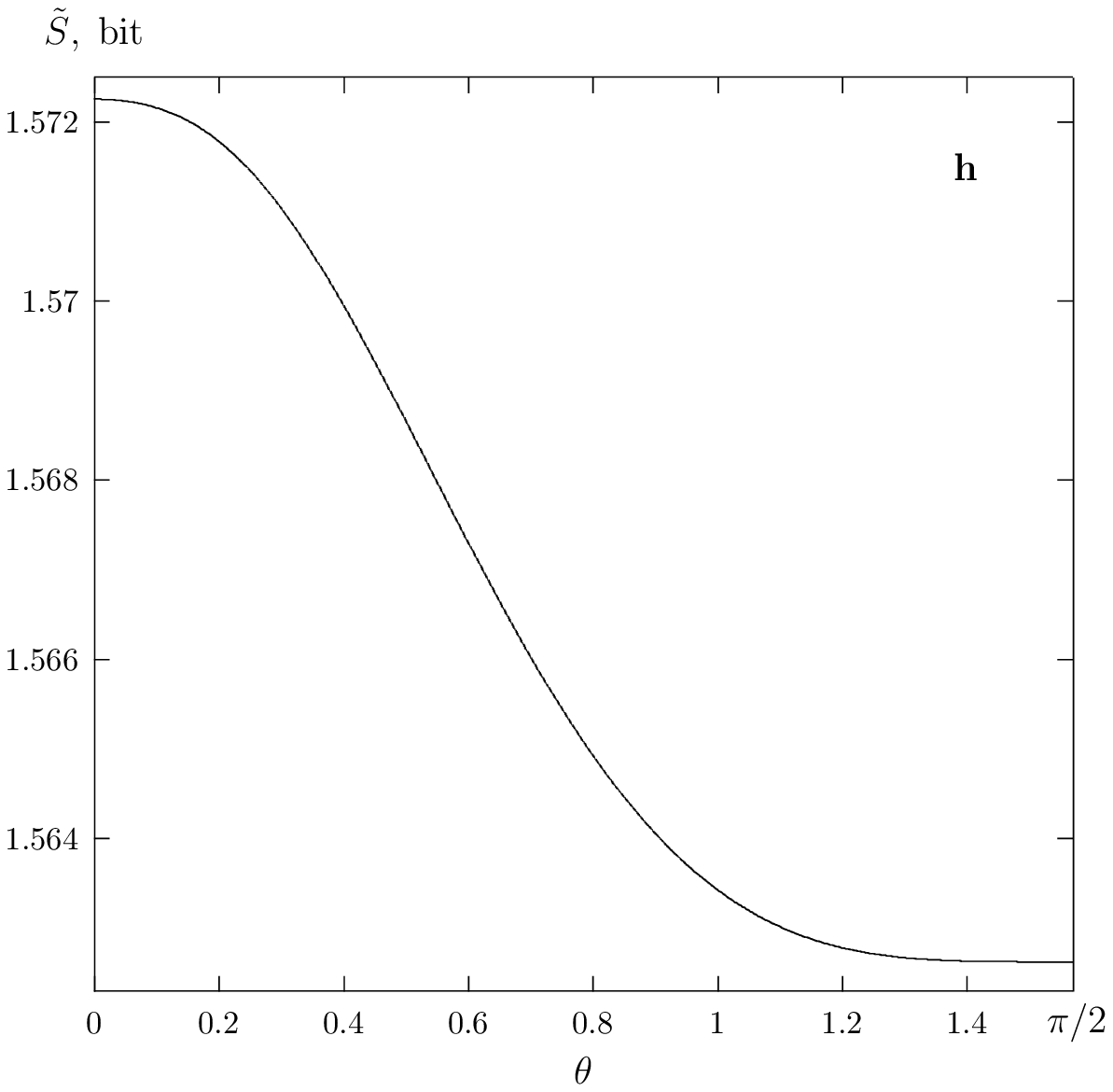,width=3.8cm}
\epsfig{file=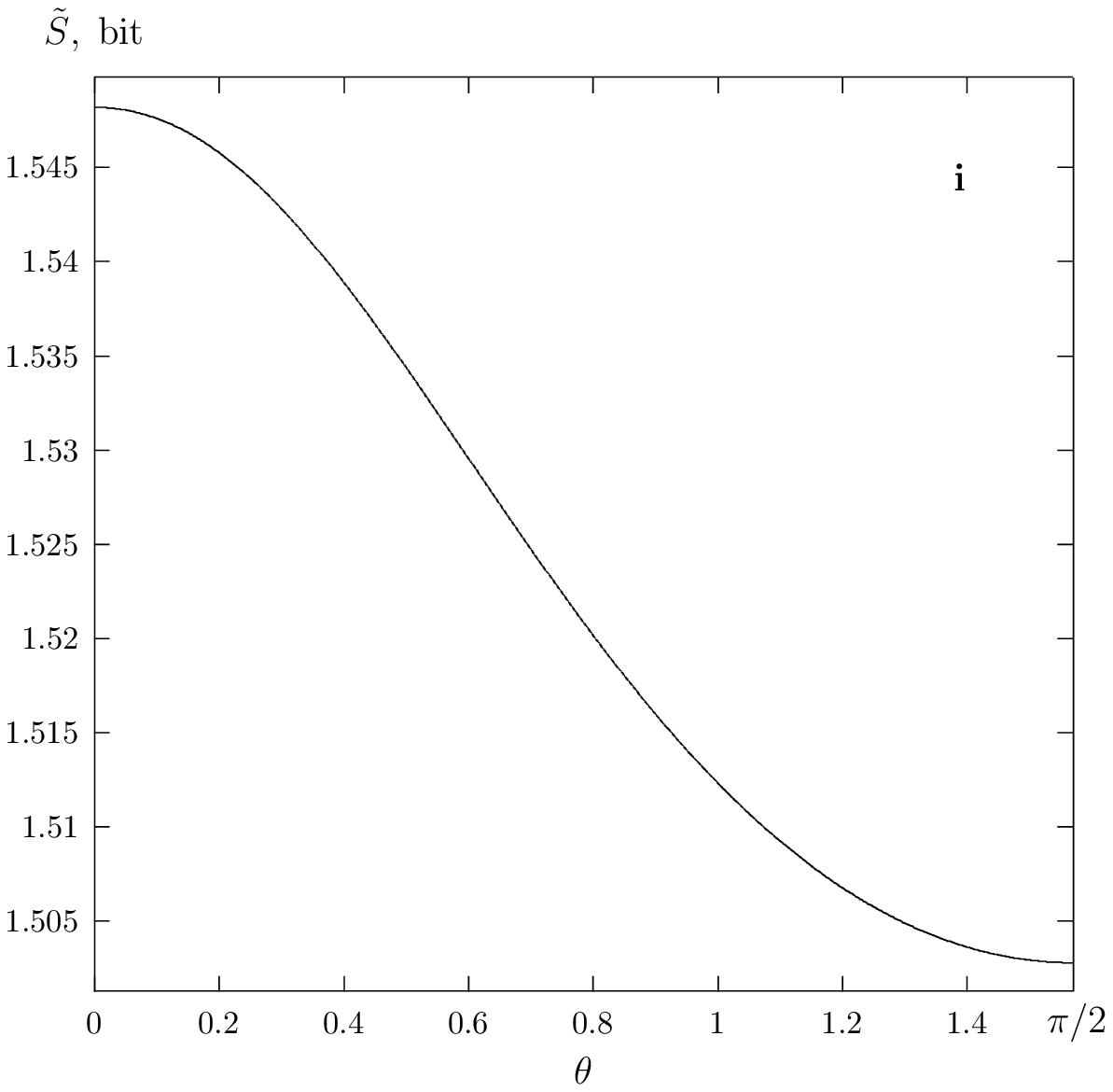,width=3.8cm}
\caption{
Post-measurement entropy $\tilde S$ vs $\theta$
by $J=-1$, $J_z=-1.5$, $B=1.9$
and $T=0.7$~(a), 0.64~(b), 0.637~(c), 0.63329~(d), 0.628~(e),
0.62378~(f), 0.61471~(g), 0.59669~(h), and 0.5~(i)
}
\label{fig:zs15mB19}       % Give a unique label
\end{center}
\end{figure*}

In order to identify each area of the plane $(T,B)$ we will again study the shapes of
the post-measurement entropy function $\tilde S(\theta)$ in all subregions.
Let us examine, for example, the shapes moving along the path $B/|J|=1.9$.
Evolution of the $\tilde S(\theta)$ curve is depicted in
Fig.~\ref{fig:zs15mB19}~(a)-(f).
%
%----------------------------------------------------------------------
%                          FIGURE 6
%----------------------------------------------------------------------
When the temperature is high enough, the curve $\tilde S(\theta)$ has the
monotonically increasing behavior as shown in Fig.~\ref{fig:zs15mB19}~(a) by $T=0.7$.
Its minimum is located at $\theta=0$ and hence this is the region $\Delta_0$
(see the right part of Fig.~\ref{fig:z15my}).

However, as the system cools, the shape of the curve begins to deform and an
inflection point with the horizontal tangent appears at $T=0.637$ [see
Fig.~\ref{fig:zs15mB19}~(b) and (c)].

With further decreasing the temperature, the dependence $\tilde S(\theta)$ becomes
bimodal in the open interval $(0,\pi/2)$.
At $T=0.63329$ we arrive at the $0^\prime$-boundary (see the right part in
Fig.~\ref{fig:z15my}) where the value of minimum reaches the value of
$\tilde S(\theta)$ at $\theta=0$.
This moment is fixed in Fig.~\ref{fig:zs15mB19}~(d).
The optimal measurement angle suddenly jumps from zero to the interior minimum angle
$\vartheta$ for any arbitrary small decreasing the temperature.

The jumps $\Delta\vartheta$ depend on the strength of magnetic field.
The values of jumps by different $B/|J|$ are collected in Table~\ref{tab:1}. 
%......................................................................
%                          TABLE 1
\begin{table}[t]
\caption{
Jumps of optimal measured angles, $\Delta\vartheta$, on the boundary $0^\prime$
for the model with $J=-1$ and $J_z/|J|=-1.5$
\upshape\upshape}
\label{tab:1}
\begin{tabular}{lll}
\hline\noalign{\smallskip}
$B/|J|$ & $T/|J|$ & $\Delta\vartheta$ \\
\noalign{\smallskip}\hline\noalign{\smallskip}
%$1.6851637$ & $0.6454108$ & $1.570782\approx90^\circ=89.999^\circ\simeq\pi/2=90^\circ$ \\
$1.6851637$ & $0.6454108$ & $1.570782\approx90^\circ$ \\
$1.7$ & $0.64533$ & $1.30773\approx74.9^\circ$ \\
$1.8$ & $0.64193$ & $0.86605\approx49.6^\circ$ \\
$1.9$ & $0.63329$ & $0.64026\approx36.7^\circ$ \\
$2.0$ & $0.61883$ & $0.48104\approx27.6^\circ$ \\
\noalign{\smallskip}\hline
\end{tabular}
\end{table}
%......................................................................
It is seen from the table that the jump is largest (equal to $\pi/2$)
at the triple point (marked in figures by black circles) and gradually disappears with
increasing $B$.

Let us go back to the path $B/|J|=1.9$ and Fig.~\ref{fig:zs15mB19}.
After crossing border $0^\prime$, the system enters region $\Delta_\vartheta$
shown on the right in Fig.~\ref{fig:z15my}.
Here the interior maximum moves to the endpoint $\theta=0$ while the interior
minimum, on the contrary, goes to the other endpoint, namely, $\theta=\pi/2$.
This situation is shown in Fig.~\ref{fig:zs15mB19}~(e).

The path intersects the dotted line at $T=0.62378$, where the values of
post-measurement entropies at both endpoints coincide, see
Fig.~\ref{fig:zs15mB19}~(f).

When the temperature reaches the value of $T=0.61471$, the inverse
process to the bifurcation is happened: the interior maximum merges with the maximum
at the endpoint $\theta=0$ and
the function $\tilde S(\theta)$ again becomes unimodal in the interval $(0,\pi/2)$.
See Fig.~\ref{fig:zs15mB19}~(g).

Then, at $T=0.59669$, the interior minimum reaches the endpoint $\theta=\pi/2$ as
shown in Fig.~\ref{fig:zs15mB19}~(h) and the optimal measurement angle continuously
changes to the stationary optimal value of $\pi/2$, see
Fig.~\ref{fig:zs15mB19}~(i).
This corresponds to the $\Delta_{\pi/2}$-region presented in the phase diagram of
one-way deficit (Fig.~\ref{fig:z15my}, right).

\begin{figure*}[t]
\begin{center}
\epsfig{file=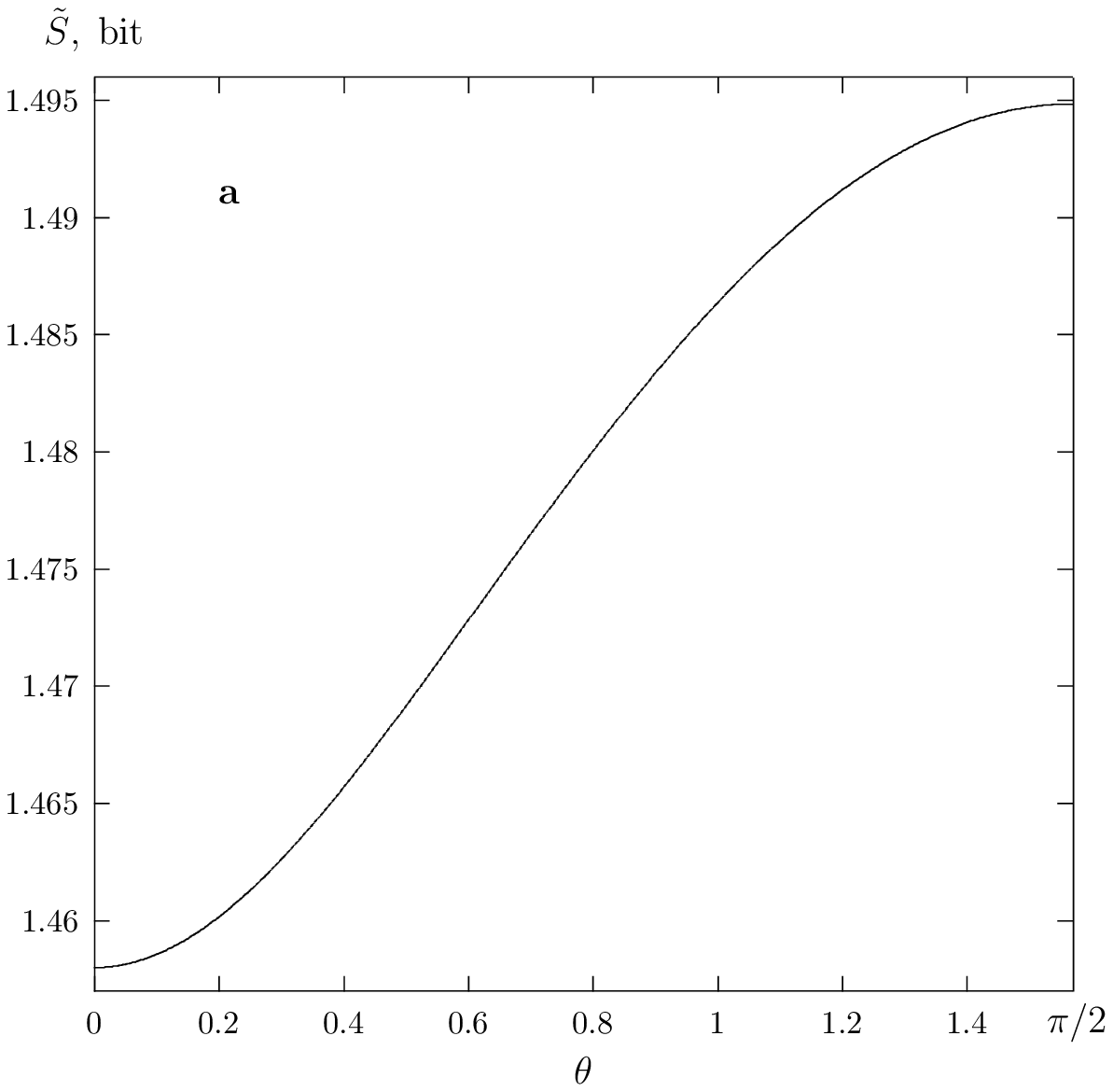,width=3.8cm}
\epsfig{file=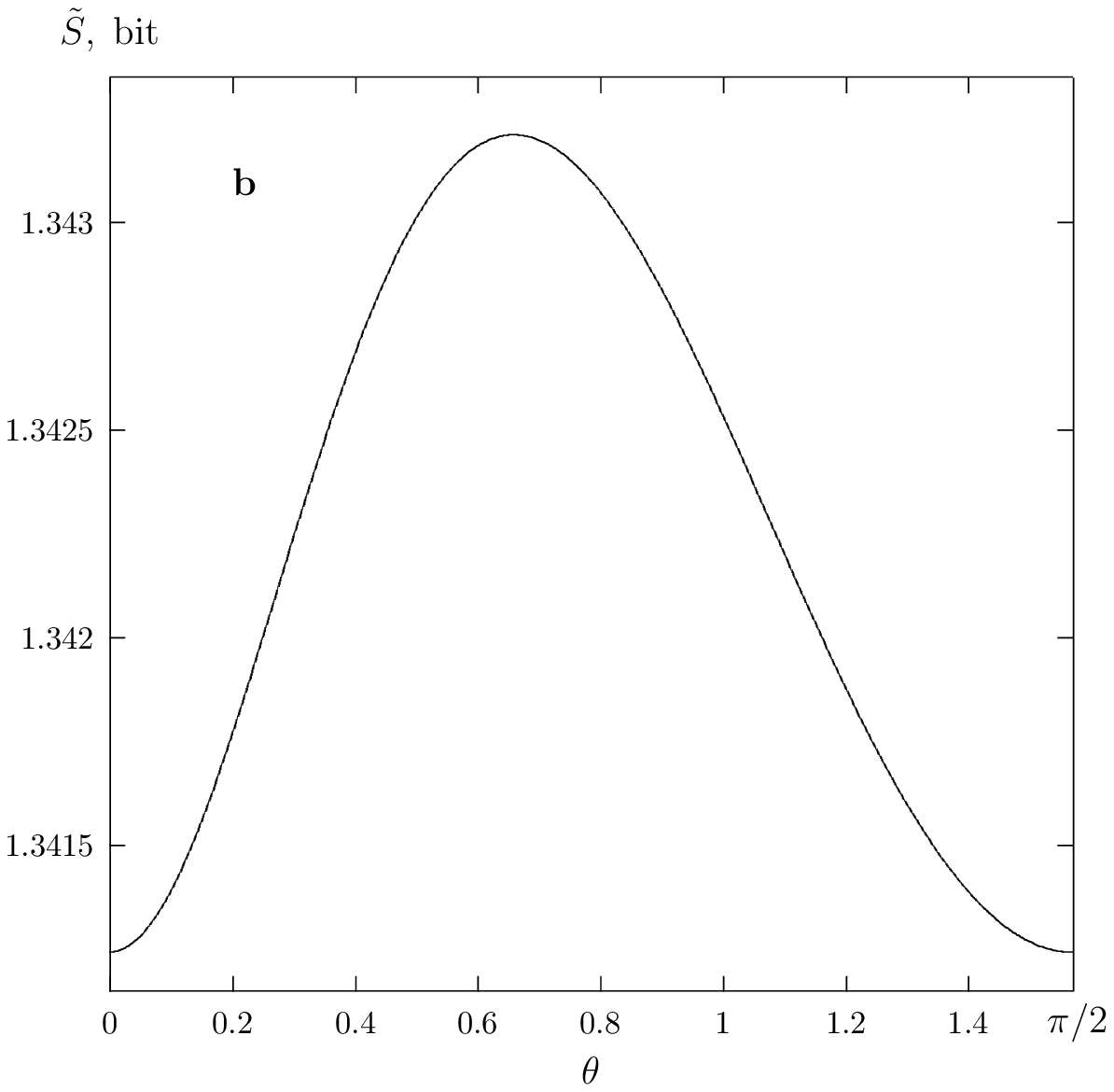,width=3.8cm}
\epsfig{file=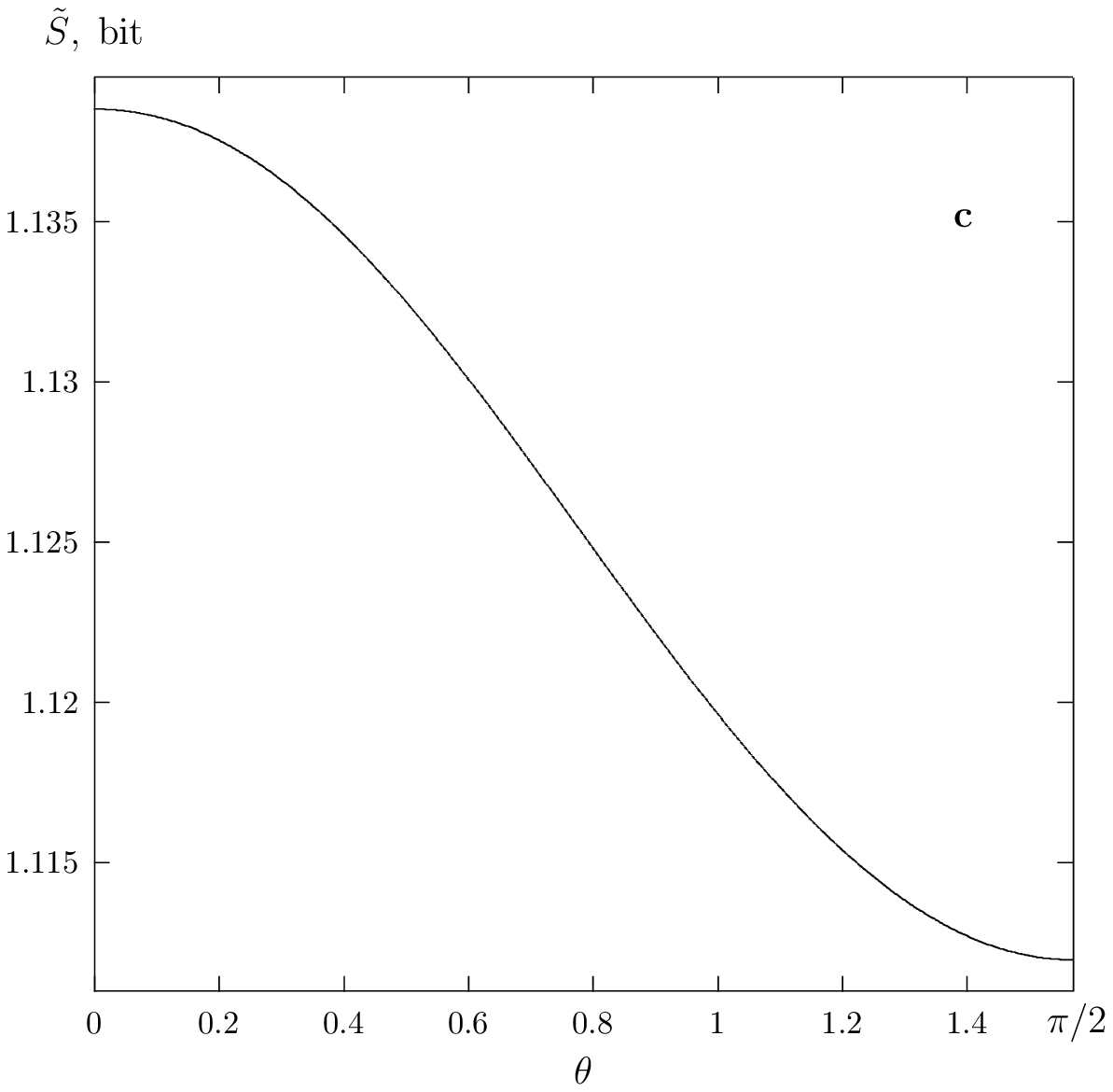,width=3.8cm}
\caption{
Post-measurement entropy $\tilde S$ vs $\theta$
by $J=-1$, $J_z=-1.5$, $B=1.2$
and $T=0.7$~(a), 0.54836~(b), 0.35~(c)
}
\label{fig:zs15mB12}       % Give a unique label
\end{center}
\end{figure*}

Let us study now another a matter of principle case, namely, the case when the
horizontal path goes below the triple point.
Take for instance the path $B/|J|=1.2$ (see Fig.~\ref{fig:z15my}, left).
The evolution of shape of the curves $\tilde S(\theta)$ by moving along this path is
depicted in Fig.~\ref{fig:zs15mB12}.
%
%......................................................................
%                          FIGURE 7
%......................................................................
When the temperature is high enough the curve $\tilde S(\theta)$ exhibits a
monotonically increasing behavior that is illustrated in Fig.~\ref{fig:zs15mB12}~(a).
The optimal measurement angle is zero and therefore this is the region $\Delta_0$.
Moving into the lower temperatures the path reaches the line 1 ($\pi/2$-boundary).
At the intersection point, a bifurcation of maximum at the
endpoint $\theta=\pi/2$ occurs and the interior maximum is born.
However, the measurement angle that minimized $\tilde S(\theta)$ stays zero.
When the dotted line is reached, the values of $\tilde S(\theta)$ at both endpoints
become equal.
This moment of equilibrium is fixed in Fig.~\ref{fig:zs15mB12}~(b).
With an arbitrary small decrease in temperature, the optimal measurement angle
discontinuously changes from zero to $\pi/2$, and the region $\Delta_{\pi/2}$ begins.
Then, at the line 0, the interior maximum annihilates at the endpoint $\theta=0$ via
the inverse bifurcation mechanism.
After that, the curve $\tilde S(\theta)$ has a monotonically decreasing shape
as drawn in Fig.~\ref{fig:zs15mB12}~(c).
This corresponds to the region $\Delta_{\pi/2}$ up to $T=0$, see
Fig.~\ref{fig:z15my}.
So, the line obeying the equation~(\ref{eq:S0S1}) serves here as the
boundary between the regions $\Delta_0$ and $\Delta_{\pi/2}$.
This boundary is shown on the right part of Fig.~\ref{fig:z15my} by the solid line and
marked by the symbol 2.
So, we complete the construction of phase diagram shown in Fig.~\ref{fig:z15my} on the
right.

\begin{figure}[t]
%\begin{figure}
\begin{center}
\epsfig{file=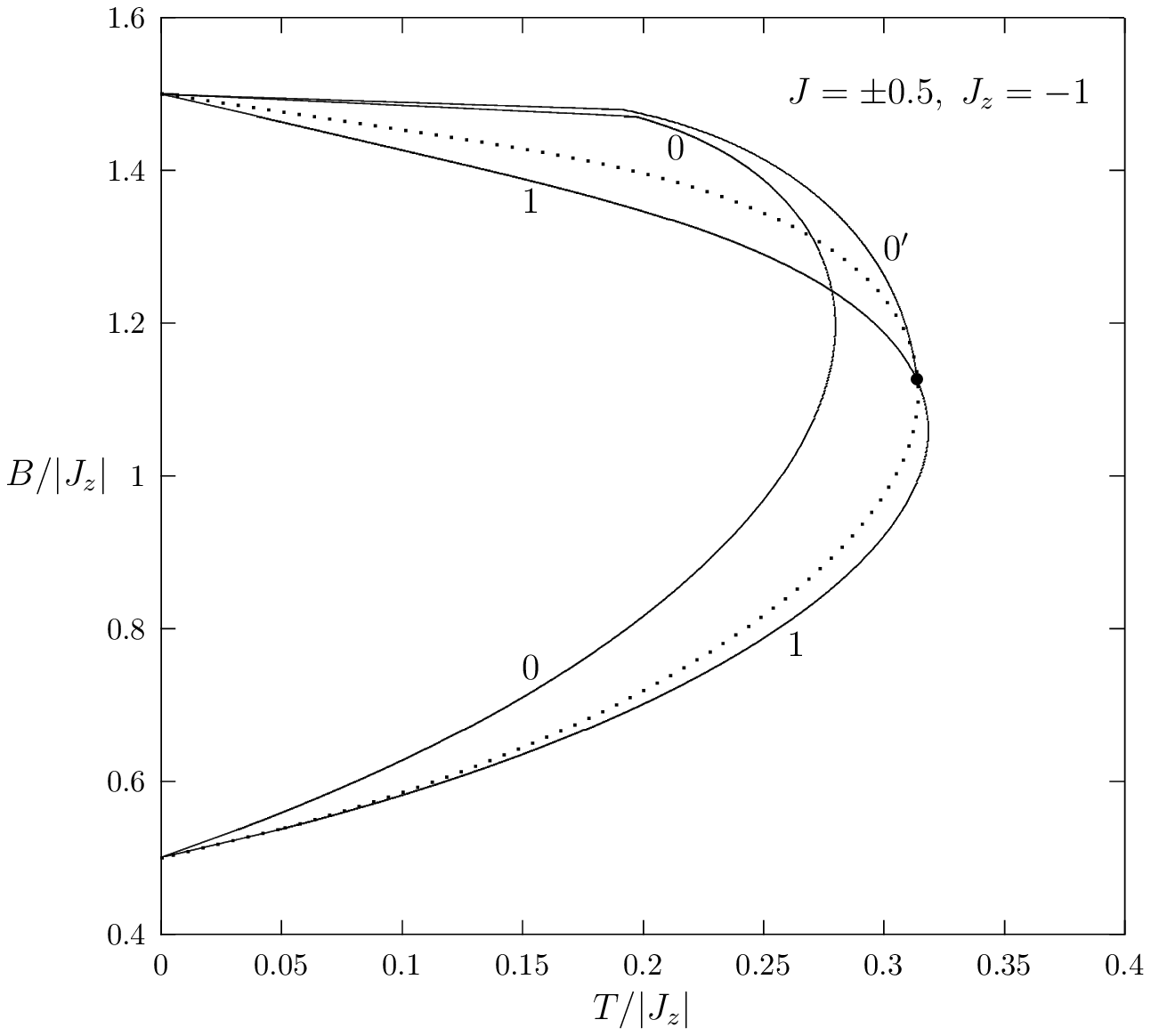,width=5.4cm}
\hspace{3mm}
\epsfig{file=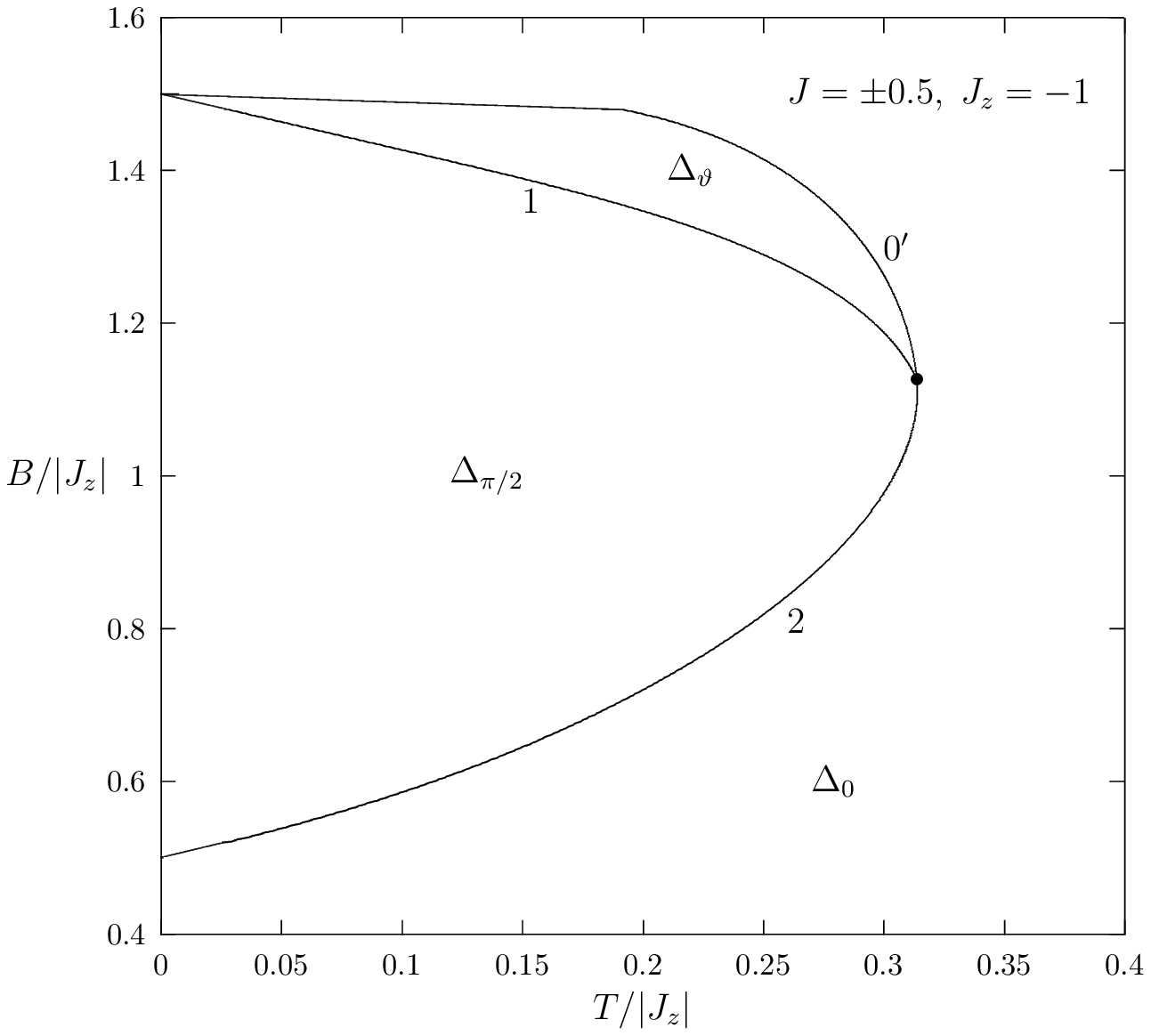,width=5.4cm}
\caption{
Lines corresponding to formal solution of the boundary equations (left) and the
phase diagram (right) for the system with $J=\pm0.5$ and $J_z=-1$.
The triple point (black circle) has coordinates $(0.313637,1.12742)$.
Normalization on $|J_z|$ is used here
}
\label{fig:z05m1m}
\end{center}
\end{figure}

\begin{figure}[t]
%\begin{figure}
\begin{center}
\epsfig{file=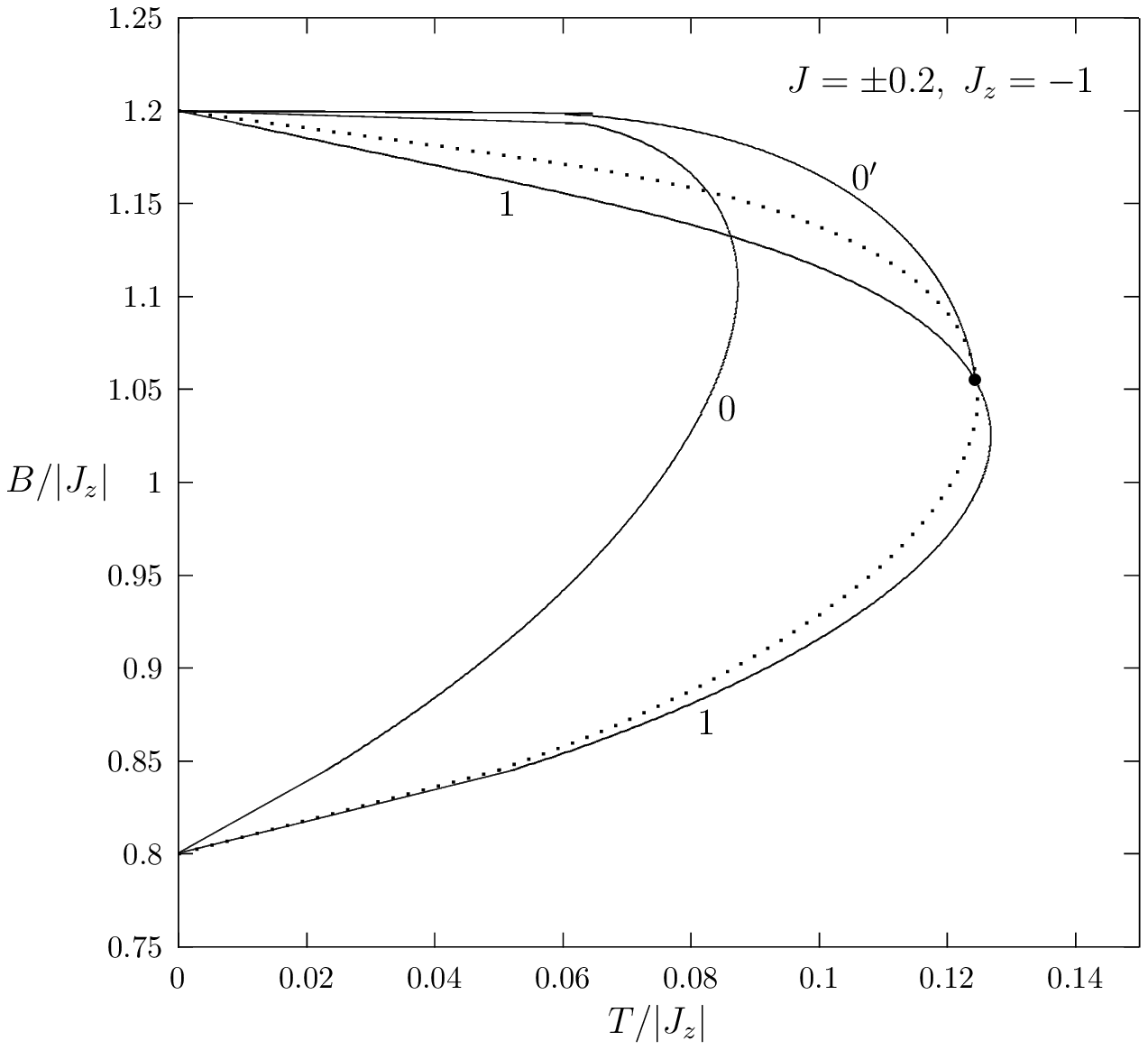,width=5.4cm}
\hspace{3mm}
\epsfig{file=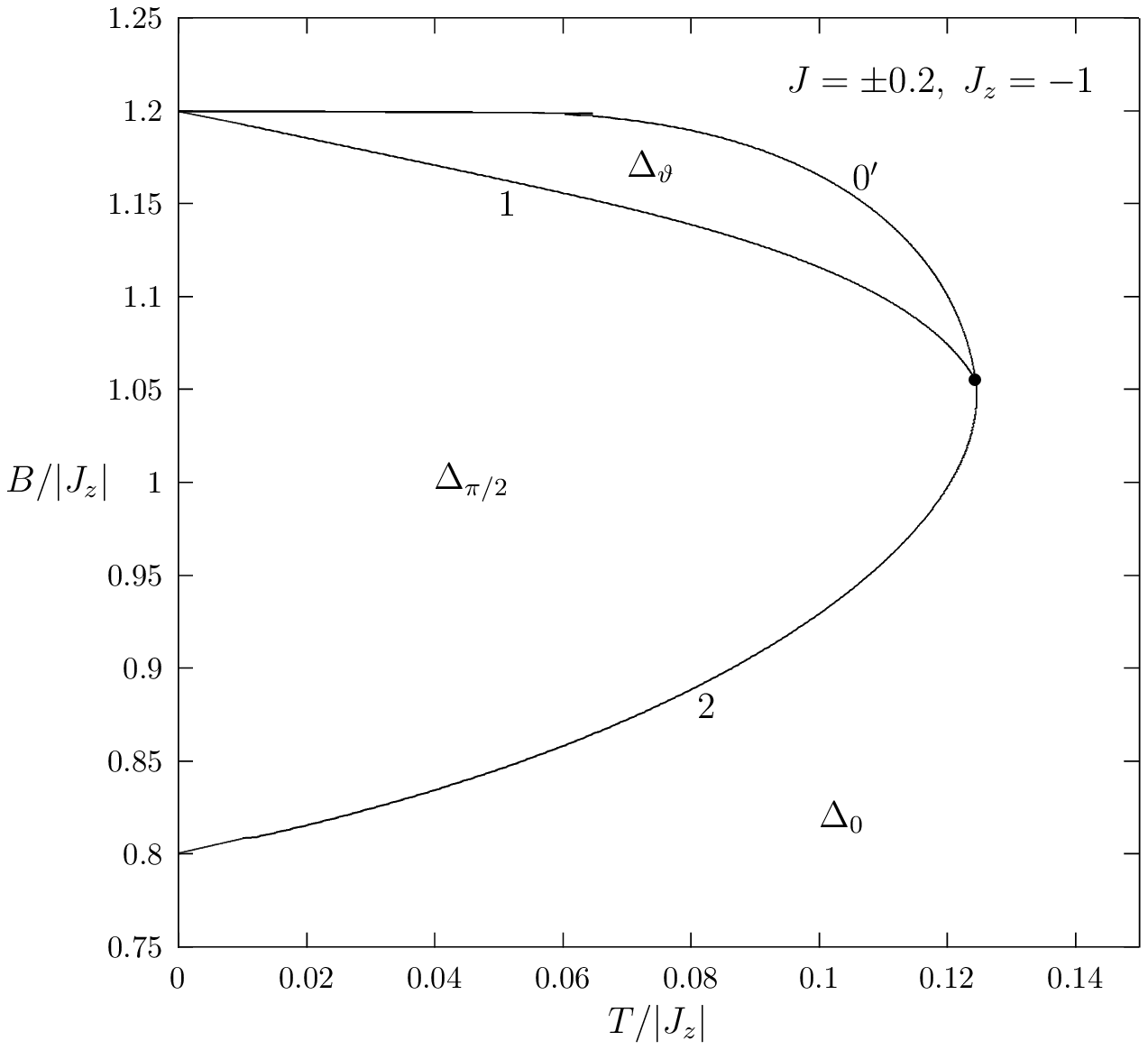,width=5.4cm}
\caption{
Lines corresponding to formal solution of the boundary equations (left) and the
phase diagram (right) for the system with $J=\pm0.2$ and $J_z=-1$.
Coordinates of the triple point are $(0.1244107,1.055204)$.
Here normalization is done on $|J_z|$
}
\label{fig:z02m1m}
\end{center}
\end{figure}

Auxiliary constructions and phase diagrams with an even more pronounced Ising's
character of interactions are presented in Figs.~\ref{fig:z05m1m} and
\ref{fig:z02m1m}.
%......................................................................
%                          FIGURE 8
%\begin{figure}[h]
%......................................................................
%
%......................................................................
%                          FIGURE 9
%\begin{figure}[h]
%......................................................................
Here the quantity $J_z$ is kept constant and the interaction $J$, on the contrary,
decreases.
This better reflects the situation in the model.
It is seen from the phase diagrams that the area of the phases $\Delta_{\pi/2}$ and
$\Delta_\vartheta$ goes to zero and vanishes at all in the limit $|J|\to0$.
As a result, the entire plane $(T,B)$ is occupied by a single phase $\Delta_0$.

This completes a construction of the phase diagrams for the parameters $J$ and $J_z$
provided $J_z<-|J|$.

%----------------------------------------------------------------------
\subsection{
The case $-|J|<J_z\le0$
}
\label{sect:Jz<1}
\begin{figure}[t]
%\begin{figure}
\begin{center}
\epsfig{file=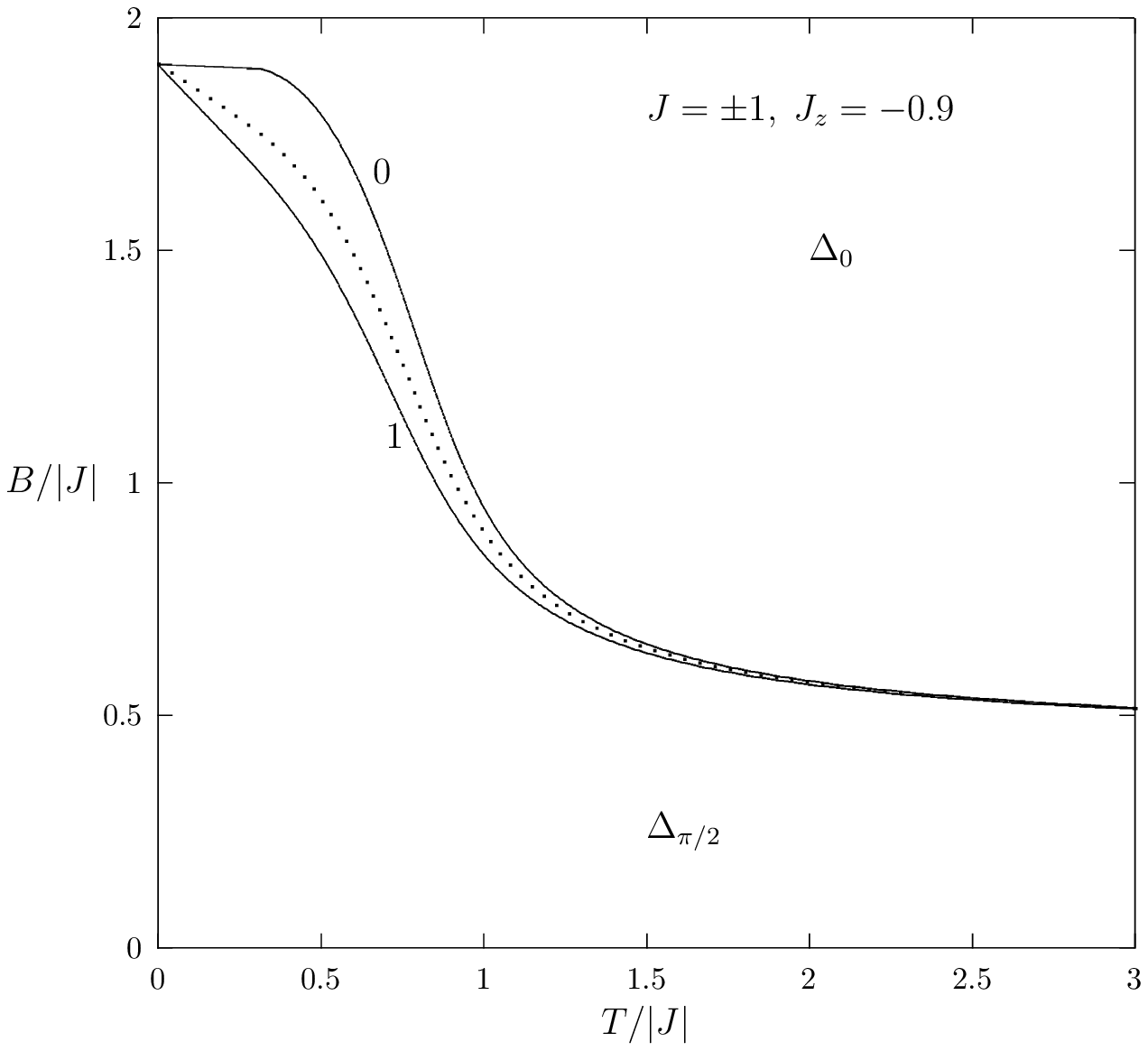,width=5.4cm}
\hspace{3mm}
\epsfig{file=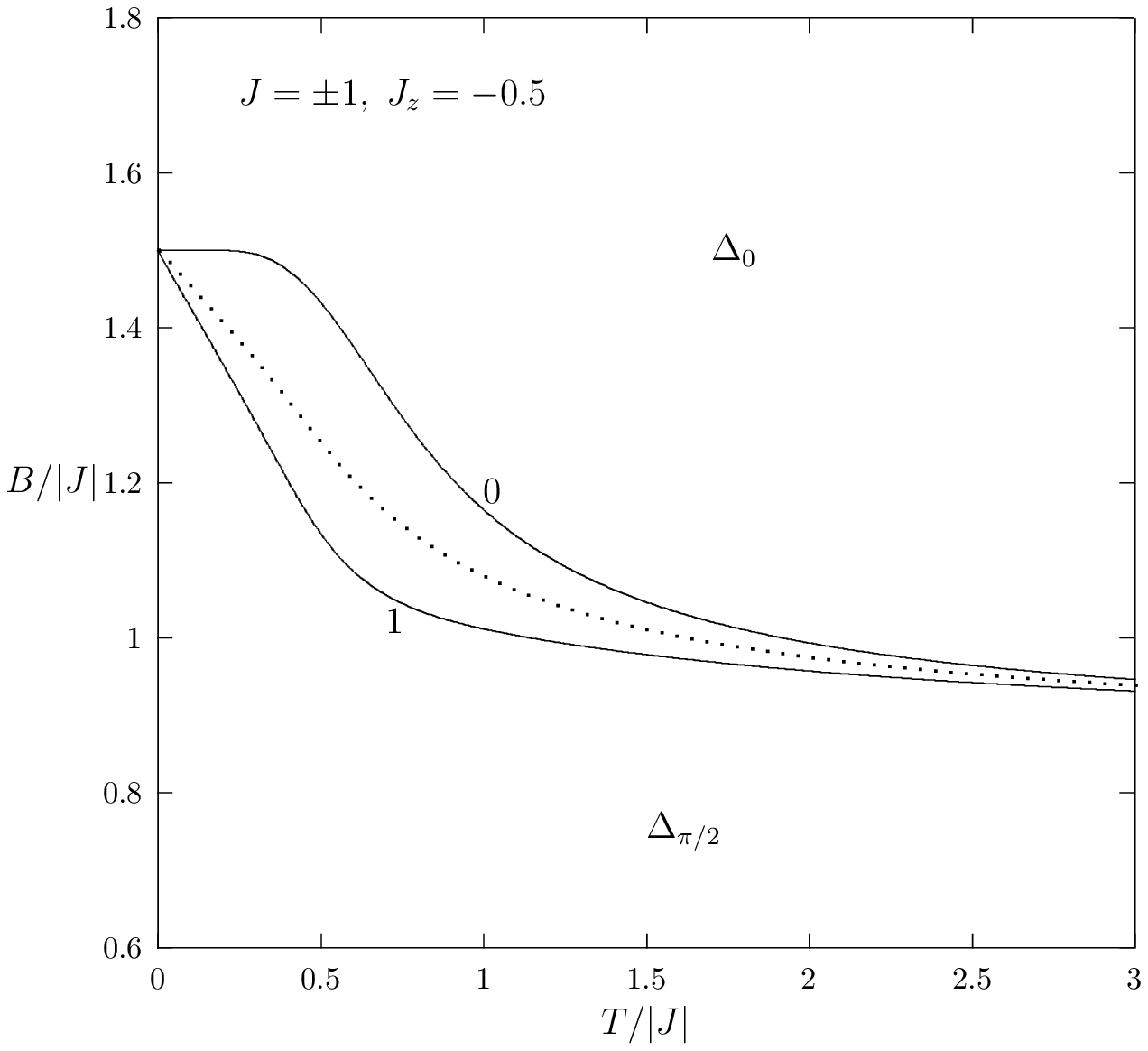,width=5.4cm}
\caption{
Phase diagram for $|J|=1$ and $J_z=-0.9$ (left) and $-0.5$ (right).
Region $\Delta_\vartheta$ lies between the boundaries 0 and 1.
Dotted line corresponds to the condition $\Delta_0=\Delta_{\pi/2}$
}
\label{fig:z09m}
\end{center}
\end{figure}

When $J_z/|J|\ge-1$, the boundary $0^\prime$ is absent or, more correctly, it
coincides the 0-boundary.
In other words, there are no jumps of optimal measurement angle $\vartheta$ on finite
steps $\Delta\vartheta>0$, but less than $\pi/2$;
the angle $\vartheta$ in the region $\Delta_\vartheta$ varies continuously from zero
to $\pi/2$.
Typical phase diagrams under question are shown in Fig.~\ref{fig:z09m}.
%......................................................................
%                          FIGURE 10
%\begin{figure}[h]
%......................................................................
The directionality of changes in curves 0 and 1 is clearly seen when the
longitudinal interaction $J_z$ tends to zero.

\begin{figure}[t]
%\begin{figure}
\begin{center}
\epsfig{file=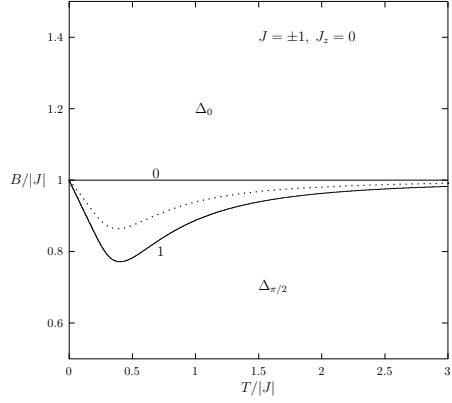,width=5.8cm}
\caption{
Phase diagram for the XX dimer $(J_z=0)$.
Region $\Delta_\vartheta$ is located between the lines 0 and 1.
Dotted line corresponds to the condition $\Delta_0=\Delta_{\pi/2}$
}
\label{fig:z0m}
\end{center}
\end{figure}

In the limit $J_z\to0$, the model is transformed into the XX one.
In accord with Eqs.~(\ref{eq:abdv}), (\ref{eq:Sii0_0}) and (\ref{eq:Sii0}), the equation for the
0-boundary by $J_z=0$ is given as
\begin{eqnarray}
   \label{eq:Sii0_0a}
		&&\sinh^2\bigg(\frac{J}{T}\bigg)\Bigg[\frac{\ln[\exp(B/T)/\cosh(J/T)]}{\exp(B/T)-\cosh(J/T)}
		+\frac{\ln[\cosh(J/T)/\exp(-B/T)]}{\cosh(J/T)-\exp(-B/T)}\Bigg]
	 \nonumber\\
%    &&=\frac{2B}{T}\sinh(B/T)-4[\cosh(B/T)-\cosh(J/T)]\ln[\cosh(J/T)].
    &&=\frac{2B}{T}\sinh\bigg(\frac{B}{T}\bigg)-4\bigg[\cosh\bigg(\frac{B}{T}\bigg)
		-\cosh\bigg(\frac{J}{T}\bigg)\bigg]\ln\!\bigg[\cosh\bigg(\frac{J}{T}\bigg)\bigg].
\end{eqnarray}
This transcendental equation can be solved analytically.
Indeed, it is easy to check that the substitution $B=|J|$ satisfies the equation
(\ref{eq:Sii0_0a}).
Thus, the 0-boundary is here the strait line $B/|J|=1$ parallel to the
abscissa axis.
The phase diagram in this limit is depicted in Fig.~\ref{fig:z0m}.
%......................................................................
%                          FIGURE 11
%\begin{figure}[h]
%......................................................................
The second boundary, line 1,
is a curve containing the interior local minimum with coordinates $(0.404,0.7716)$.
The fidelity between quantum states at this point and the nearest one on the 
0-boundary, i.e., at point $(0.404,1)$ equals $F=0.97994$ ($\approx97.99\%$).
This value characterizes the width of $\Delta_\vartheta$-region.
The region near this minimum is broad enough and therefore one can try to reach it
in an experiment.

%----------------------------------------------------------------------
\subsection{
The case $0<J_z\le|J|$
}
\label{sect:Jz<1b}
%
%......................................................................
%                          FIGURE 12
%\begin{figure}[h]
\begin{figure}[t]
%\begin{figure}[b]
%\begin{figure}
\begin{center}
\epsfig{file=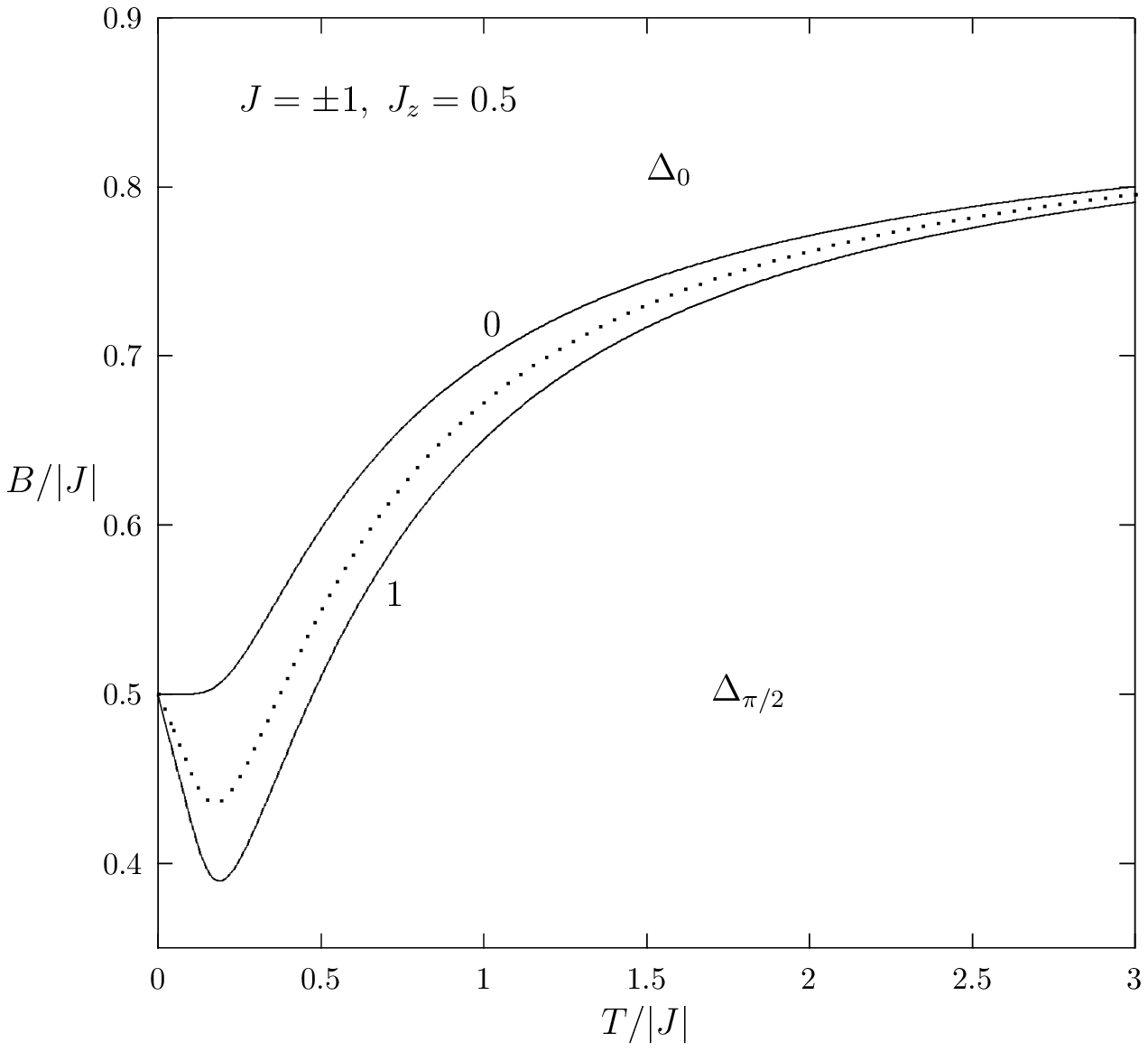,width=5.4cm}
\hspace{3mm}
\epsfig{file=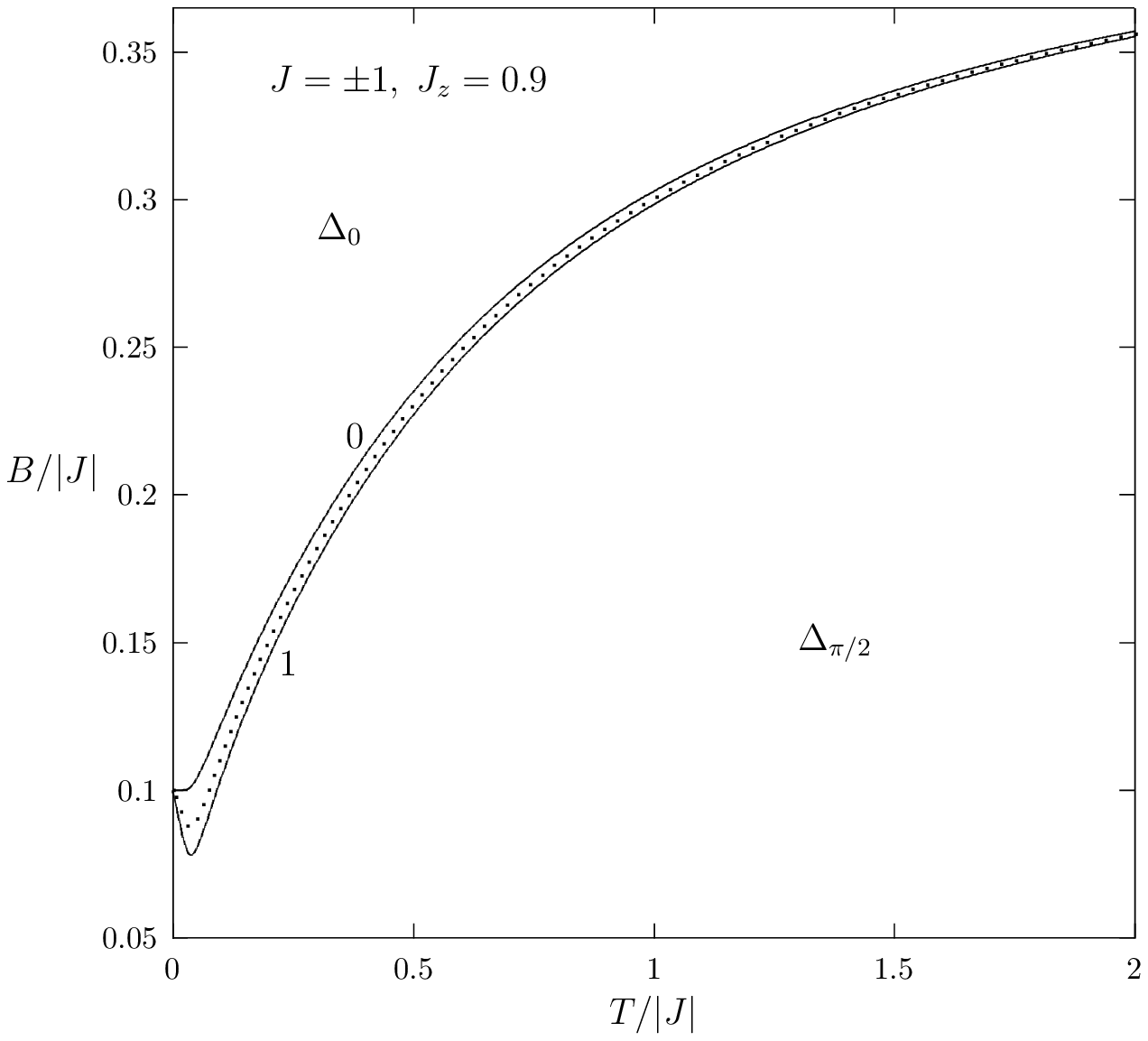,width=5.4cm}
\caption{
Phase diagram for $|J|=1$ and $J_z=0.5$ (left) and $0.9$ (right).
The region $\Delta_\vartheta$ lies between the boundaries 0 and 1.
Dotted line corresponds to the condition $\Delta_0=\Delta_{\pi/2}$
}
\label{fig:z05p}
\end{center}
\end{figure}
%......................................................................
So when $J_z=0$, the phase diagram looks as shown in Fig.~\ref{fig:z0m}.
Let us continue to increase the value of interaction $J_z$.
Phase diagrams at $J_z/|J|=0.5$ and 0.9 are presented in Fig.~\ref{fig:z05p}.
One can see that the 0- and $\pi/2$-boundaries drop down to the abscissa axis when the
quantity $J_z/|J|$ tends to unity.
In the limit $Jz/|J|\to1$, the regions $\Delta_{\pi/2}$ and $\Delta_\vartheta$
disappear completely and the phase $\Delta_0$ occupies the entire area of plane
$(T,B)$.

%----------------------------------------------------------------------
\subsection{
The case $J_z>|J|$
}
\label{sect:Jz>1}
When the coupling $J_z$ is grater than $|J|$, the system belongs to
the Ising-like ferromagnetic type.
The $z$-component of the spin interaction dominates.
As the analysis shows, the phase diagram consists here of one fraction $\Delta_0$
at any temperatures $T$ and strengths of the field $B$.

It is interesting to compare this result with the behavior of quantum discord.
In contrast to the one-way deficit, the quantum discord exhibits different behavior
for the same choice of coupling constants.
Indeed, as shown in Fig.~4 of Ref.~\cite{Y14}
 for $J=1$ and $J_z=1.02$,
the phase diagram of the quantum
discord consists of three regions corresponding to the optimal measurement angles
zero, $\pi/2$, and an angle in the intermediate interval between zero and $\pi/2$
(the region with the interior optimal measurement angle).

Such a discrepancy indicates the imperfectness of existing concepts of quantum
correlations.
We unfortunately have to admit that we do not know for sure what measure of quantum
correlation is true.
Only further progress in the field of theory and practice of quantum correlations will
help resolve the contradictions existing today.

%======================================================================
\section{Summary and conclusions}
\label{sec:Concl}
In this paper, we have performed analysis and constructed the temperature-field phase
diagrams of one-way quantum work deficit for the two-qubit XXZ model in a uniform
magnetic field.
The set of figures presented in Sect.~\ref{sec:PhDs}
(first of all, Figs.~\ref{fig:z1m1a}, \ref{fig:z15my}, \ref{fig:z09m},
\ref{fig:z0m}, and \ref{fig:z05p})
gives a complete picture of the qualitatively different types of $(T,B)$-diagrams in
the entire range of values for the parameter $J_z$, i.e., in the three-dimensional
space $(T,B,J_z)$.

We have also described the special properties of post-measurement entropy that are
important in planning and performing an experiment.
The diagrams allow to predict where the quantum correlation will experience the
different features due to jumps of optimal measurement angle.
Thus, the presented diagrams can serve as a kind of routings useful for experimenters
in their work.

In addition, we have established a connection between the cause of the appearance of
boundaries separated different fractures of the quantum correlation and Landau's
approach to the problem of phase transitions as well as the mathematical theory of
catastrophes.

Further development of the presented results is associated with the consideration of
more general spin systems.
However, this will require consideration of multidimensional spaces, which will
certainly make the problem much more difficult.
One can rely a hope for virtual reality methods which are developed now very
rapidly.

%======================================================================
%\section*{Acknowledgment}
\vspace{-10mm}
\section*{}
{\bf Acknowledgments}\ I am grateful to A.~I.~Zenchuk for his help.
This work was performed as a part of the state task, State Registration
No.~0089-2019-0002.

%======================================================================
%\clearpage
%\newpage

%======================================================================


\begin{thebibliography}{99}

\bibitem{MBCPV12}
Modi,~K., Brodutch,~A., Cable,~H., Paterek,~T., Vedral,~V.:
The classical-quantum boundary for correlations: discord and related measures.
Rev. Mod. Phys. {\bf 84}, 1655 (2012)

\bibitem{Str15}
Streltsov,~A.:
Quantum correlations beyond entanglement and their role in quantum information
theory.
%SpringerBriefs in Physics.
Springer, Berlin (2015)
%, arXiv:1411.3208v1~[quant-ph]

\bibitem{ABC16}
Adesso,~G., Bromley,~T.R., Cianciaruso,~M.:
Measures and applications of quantum correlations.
J. Phys. A: Math. Theor. {\bf 49}, 473001 (2016)

\bibitem{FPA17}
Lectures on general quantum correlations and their applications.
Eds: Fanchini,~F.F., Soares-Pinto,~D.O., Adesso,~G.
Springer, Berlin (2017)

\bibitem{BDSRSS18}
Bera,~A., Das,~T., Sadhukhan,~D., Roy,~S.S., Sen(De),~A., Sen,~U.:
Quantum discord and its allies: a review of recent progress.
Rep. Prog. Phys. {\bf 81}, 024001 (2018)
%, arXiv:1703.10542v1 [quant-ph]

\bibitem{FLS64}
Feynman,~R.P., Leighton,~R.B., Sands, M.:
The Feynman lectures on physics.
Addison-Wesley, Reading, Mass. (1964, second printing). - Sect.~38-6

\bibitem{H25}
Heisenberg,~W.:
$\ddot {\rm U}$ber quantentheoretische Umdeutung kinematischer und mechanischer
Beziehungen.
Zs. Phys. {\bf 33}, 879 (1925)

\bibitem{W14}
Wolff,~J.:
Heisenberg's observability principle.
{\em In}: Studies in history and philosophy of science. Part~B. Studies in history and
philosophy of modern physics {\bf 45}, 19 (2014)

\bibitem{Z00}
Zurek,~W.H.:
Einselection and decoherence from an information theory perspective.
Ann. Phys. (Leipzig) {\bf 9}, 855 (2000)

\bibitem{OZ02}
Ollivier,~H., Zurek,~W.H.:
Quantum discord: a measure of the quantumness of correlations.
Phys. Rev. Lett. {\bf 88}, 017901 (2002)

\bibitem{HV01}
Henderson,~L., Vedral,~V.:
Classical, quantum and total correlations.
J. Phys. A: Math. Gen. \textbf{34}, 6899 (2001)

\bibitem{V17}
Vedral,~V.:
Foundations of quantum discord.
{\em In}:
Lectures on general quantum correlations and their applications.
Eds: Fanchini,~F.F., Soares-Pinto,~D.O., Adesso,~G.
Springer, Berlin (2017)

\bibitem{L73}
Lindblad,~G.:
Entropy, information and quantum measurements.
Commun. Math. Phys. \textbf{33}, 305 (1973)

\bibitem{OHHH02}
Oppenheim,~J., Horodecki,~M., Horodecki,~P., Horodecki,~R.:
Thermodynamical approach to quantifying quantum correlations.
Phys. Rev. Lett. {\bf 89}, 180402 (2002)

\bibitem{HHHHOSS03}
Horodecki,~M., Horodecki,~K., Horodecki,~P., Horodecki,~R., Oppenheim,~J., Sen(De),
A., Sen, U.:
Local information as a resource in distributed quantum systems.
Phys. Rev. Lett. {\bf 90}, 100402 (2003)

\bibitem{HHHOSSS05}
Horodecki,~M., Horodecki,~P., Horodecki,~R., Oppenheim,~J., Sen(De), A., Sen, U.,
Synak-Radtke, B.:
Local versus nonlocal information in quantum-information theory: Formalism and
phenomena.
Phys. Rev. A {\bf 71}, 062307 (2005)

\bibitem{YF16}
Ye,~B.-L., Fei,~S.-M.:
A note on one-way quantum deficit and quantum discord.
Quantum Inf. Process. {\bf 15}, 279 (2016)

\bibitem{F65}
Feynman,~R.:
The character of physical law.
MIT, Cambridge, Mass. (1985, twelfth printing). - Sect.~7 Seeking new laws, p.~164

\bibitem{LXSW10}
Lu,~X.-M., Xi,~Z., Sun,~Z., Wang,~X.:
Geometric mesure of quantum discord under decoherence.
Quantum Inf. Comput. {\bf 10}, 0994 (2010)

\bibitem{CRC10}
Ciliberti,~L., Rossignoli,~R., Canosa,~N.:
Quantum discord in finite $XY$ chains.
Phys. Rev. A {\bf 82}, 042316 (2010)

\bibitem{VR12}
Vinjanampathy,~S., Rau,~A.R.P.:
Quantum discord for qubit-qudit systems.
J. Phys. A: Math. Theor. {\bf 45}, 095303 (2012)

\bibitem{YWF16}
Ye,~B.-L., Wang,~Y.-K., Fei,~S.-M.:
One-way quantum deficit and decoherence for two-qubit $X$ states.
Int. J. Theor. Phys. {\bf 55}, 2237 (2016)

\bibitem{Y18}
Yurischev,~M.A.:
Bimodal behavior of post-measured entropy and one-way quantum deficit
for two-qubit X states.
Quantum Inf. Process. {\bf 17}:6 (2018)


\bibitem{Y19}
Yurischev,~M.A.:
Phase diagram for the one-way quantum deficit of two-qubit X states.
Quantum Inf. Process. {\bf 18}:124 (2019)
%arXiv:1804.03755v1 [quant-ph]

\bibitem{LMXW11}
Lu,~X.-M., Ma,~J., Xi,~Z., Wang,~X.:
Optimal measurements to access classical correlations of two-qubit states.
Phys. Rev. A {\bf 83}, 012327 (2011)

\bibitem{CZYYO11}
Chen,~Q., Zhang,~C., Yu,~S., Yi,~X.X., Oh,~C.H.:
Quantum discord of two-qubit $X$ states.
Phys. Rev. A {\bf 84}, 042313 (2011)

\bibitem{H13}
Huang,~Y.:
Quantum discord for two-qubit $X$ states:
analytical formula with very small worst-case error.
Phys. Rev. A {\bf 88}, 014302 (2013)

\bibitem{Y14}
Yurischev,~M.A.:,
Quantum discord for general X and CS states: a piecewise-analytical-numerical formula,
arXiv:1404.5735v1 [quant-ph]

\bibitem{Y14a}
Yurishchev,~M.A.:,
NMR dynamics of quantum discord for spin-carrying gas molecules in a closed nanopore.
J. Exp. Theor. Phys. {\bf 119}, 828 (2014); arXiv:1503.03316v1~[quant-ph]

\bibitem{Y15}
Yurischev,~M.A.:
On the quantum discord of general $X$ states.
Quantum Inf. Process. {\bf 14}, 3399 (2015)

\bibitem{Y17}
Yurischev,~M.A.:
Extremal properties of conditional entropy and quantum discord for XXZ, symmetric
quantum states.
Quantum Inf. Process. {\bf 16}:249 (2017)
%; arXiv:1702.03728v3~[quant-ph]

\bibitem{CCR15}
Canosa,~N., Ciliberti,~L., Rossignoli,~R.:
Quantum discord and information deficit in spin chains.
Entropy {\bf 17}, 1634 (2015)
(Special issue: Quantum computation and information: Multi-particle aspects)

\bibitem{K31}
Klein,~O.:
Zur quantenmechanischen Begr$\ddot {\rm u}$ndung des zweiten Hauptsatzes der
W$\ddot {\rm a}$rmelehre.
Zs. Phys. {\bf 72}, 767 (1931)

\bibitem{NC10}
Nielsen,~M.A., Chuang,~I.L.:
Quantum computation and quantum information.
10th anniversary edition.
Cambridge University Press, Cambridge (2010), Sect.~11.3.3

\bibitem{F72}
Fain,~V.M.:
Quantum radiophysics. Vol.~1. Photons and nonlinear media.
Second edition.
Sovetskoe Radio, Moscow (1972) [in Russian]

\bibitem{D58}
Dirac,~P.A.M.:
The principles of quantum mechanics.
Fourth edition.
Clarendon, Oxford (1958), Sect.~II.10

\bibitem{JKMW01}
James,~D.F.V., Kwiat,~P.G., Munro,~W.J., White,~A.G.:
Measurement of qubits.
Phys. Rev. A {\bf 64}, 052312 (2001)

\bibitem{Y19a}
Yurischev,~M.A.:
Jumps of optimal measurement angle and fractures on the curves of quantum correlation
functions.
Proc. of SPIE {\bf 11022}, 1102221 (2019)

\bibitem{MGY19}
Moreva,~E.V., Gramegna,~M., Yurischev,~M.A.:
On the possibility to detect quantum correlation regions with the variable optimal
measurement angle.
Eur. Phys. J. D {\bf 73}:68 (2019),
topical issue ``Quantum correlations''

\bibitem{LL_StPh}
Landau,~L.D., Lifshitz,~E.M.:
Statistical physics. Part~1.
Fizmatlit, Moscow (2005) [in Russian],
Pergamon, Oxford (1980) [in English]

\bibitem{A92}
Arnold, V.I.:
Catastrophe theory.
Springer, Berlin (1992), Sect.~10

\end{thebibliography}
\end{document}